\documentclass{aa}

\usepackage[varg]{txfonts}
\usepackage{graphicx}
\usepackage{array}
\usepackage{lscape}                                
\usepackage{txfonts}
\usepackage{caption}
\usepackage[normalem]{ulem}
\usepackage[dvipsnames]{xcolor}

\usepackage{longtable}
\usepackage{color}
\usepackage{supertabular}
\usepackage[draft]{hyperref}
\usepackage[flushleft]{threeparttable}
\usepackage{multirow}
\usepackage{lscape}
\usepackage{placeins}

\newcommand{\kms} {km\,s$^{-1}$}

\newcommand{\vsini} {$v$\,sin\,$i$}
\newcommand{\vini} {$v_{\rm ini}$}
\newcommand{\veq} {$v_{\rm eq}$}
\newcommand{\vcrit} {$v_{\rm crit}$}

\newcommand{\vmacro} {$v_{\rm mac}$}
\newcommand{\vrad} {$v_{\rm rad}$}
\newcommand{\Teff} {$T_{\rm eff}$}
\newcommand{\grav} {log\,{\em $g$}}
\newcommand{\gravt} {log\,{\em $g_{\rm corr}$}}

\newcommand{\micro} {$\xi_{\rm t}$}
\newcommand{\helio} {Y$_{\rm He}$}
\newcommand{\fastwind} {{\sc fastwind}}
\newcommand{\ioni}[2]{{#1\,\sc{#2}}}

\newcommand{\msol}{$M_{\odot}$}
\newcommand{\Lsp} {log\,($\mathcal{L}$/$\mathcal{L_{\odot}}$)}

\newcommand{\vobs}{$v_{\rm obs}$}

\begin{document}

\title{The IACOB project}

\subtitle{VII. The rotational properties of Galactic massive O-type stars revisited
\thanks{Table~\ref{tableValues} is only available at the CDS via anonymous ftp to cdsarc.u-strasbg.fr (130.79.128.5)
or via http://cdsweb.u-strasbg.fr/cgi-bin/qcat?J/A+A/}
} 

\author{G.~Holgado\inst{1,2,3}, S.~Sim\'on-D\'iaz\inst{1,2}, A.~Herrero\inst{1,2}, R.~H.~Barb\'a\inst{4,\dag}}

\institute{Instituto de Astrof\'isica de Canarias, E-38200 La Laguna, Tenerife, Spain.
             \and
             Departamento de Astrof\'isica, Universidad de La Laguna, E-38205 La Laguna, Tenerife, Spain.
             \and 
             Centro de Astrobiologia (CAB), CSIC-INTA, Camino Bajo del Castillo s/n, 28692, Villanueva de la Cañada, Madrid, Spain
            \and
             Departamento de F\'isica y Astronom\'ia, Universidad de La Serena, Avenida Juan Cisternas 1200, La Serena, Chile.
             }
                   
\offprints{gholgado@cab.inta-csic.es \\ \dag\ deceased.}

\date{Date}

\titlerunning{Rotational velocities of Galactic O-type stars}
\authorrunning{Holgado et al.}


%
\abstract
{Stellar rotation is of key importance in the formation process, the evolution, and the final fate of massive stars.
}
{
We perform a reassessment of the empirical rotational properties of Galactic massive O-type stars using the results from a detailed analysis of ground-based multi-epoch optical spectra obtained in the framework of the IACOB \& OWN surveys. 
}
{Using high-quality optical spectroscopy, we established the velocity distribution for a sample of 285 apparently single and single-line spectroscopic binary (SB1) Galactic O-type stars.
We also made use of the rest of the parameters from the quantitative spectroscopic analysis presented in prior IACOB papers (mainly \Teff, \grav, and multiplicity) to study the \vsini\ behavior and evolution from the comparison of subsamples in different regions of the spectroscopic Hertzsprung-Rusell diagram (sHRD). 
Our results are compared to the main predictions -- regarding current and initial rotational velocities -- of two sets of well-established evolutionary models for single stars, as well as from population synthesis simulations of massive stars that include binary interaction.  
}
{We reassess the known bimodal nature of the \vsini\ distribution, and find a non-negligible difference between the \vsini\ distribution of single and SB1 stars. 
We provide empirical evidence supporting the proposed scenario that the tail of fast rotators is mainly produced by binary interactions.
Stars with extreme rotation ($>$300~\kms) appear as single stars that are located in the lower zone of the sHRD.
We notice little rotational braking during the main sequence, a braking effect independent of mass (and wind strength). 
The rotation rates of the youngest observed stars lean to an empirical initial velocity distribution with $\lessapprox$20\% of critical velocity.
Lastly,  a limit in \vsini\ detection below 40-50~\kms\ seems to persist,
especially in the upper part of the sHRD, possibly associated with the effect of microturbulence in the measurement methodology used.
}
{}
%
\keywords{Stars: early-type -- Stars: rotation -- Techniques: spectroscopic -- Catalogs -- The Galaxy}

%
\maketitle
%
%

\section{Introduction}\label{sectionIntro}

Stellar rotation affects the global physical characteristics and the evolution of massive stars in many different ways \citep[][]{Maeder2000,Langer2012}. As a massive star evolves, it is expected that its surface rotational velocity will be modified due to different processes. Generally speaking, in single stars the equatorial velocity always decreases monotonically, and the net braking rate of the stellar surface mostly depends on the combined effect of angular momentum losses because of stellar winds and the efficiency of internal transport of angular momentum from the core to the envelope. 
In binary systems -- especially in cases where the two components are close enough to interact at some point in their lives -- the situation becomes a bit more complex. There are several additional effects that do not operate in single stars (such as tides and mass transfer events) that are capable of both decelerating or accelerating each of the individual components \citep[][and references therein]{deMink2013}. 

In order to advance our understanding of the global rotational properties of massive stars -- as well as the impact of this important parameter on their internal structure and evolution --, numerous studies have obtained and analyzed spectroscopic data~sets of difference size and quality \citep[][]{Conti1977,Penny1996,Howarth1997,Huang2006,Huang2006a,Dufton2006,Daflon2007,Hunter2008,Penny2009,Fraser2010,Braganca2012,Simon-Diaz2014,Ramirez-Agudelo2013,Ramirez-Agudelo2015}. Some of them have also compared the gathered empirical information with the general predictions of evolutionary models and population synthesis computations that include rotation \citep[][]{Simon-Diaz2010,Sundqvist2013,Granada2014,Martins2017,Markova2014,Markova2018,Keszthelyi2017,Keszthelyi2020}. Others have used the obtained distributions of projected rotational velocities (\vsini) to establish empirical constraints to our theories of massive star formation \citep[][]{Wolff2006,Mokiem2006,Huang2010,Rosen2012}.

Following \citep{Holgado2018, Holgado2020}, this paper is a continuation of a series devoted to the study of a sample of more than 400 Galactic O-type stars, for which the IACOB and OWN surveys \citep[last described in][]{SimonDiaz2020, Barba2017} have been gathering high-quality, multi-epoch spectroscopic observations over the last few decades. In this paper, we focus on presenting a re-evaluation of our empirical knowledge about the global spin-rate properties of the Galactic main sequence stars with masses in the range $\sim$15\,--\,80~\msol.

To this aim, we used the outcome of a homogeneous (and objective) quantitative spectroscopic analysis, performed with a battery of semi-automated tools based on state-of-the-art techniques \citep{Simon-Diaz2011, Simon-Diaz2014}, and the stellar atmosphere code {\sc fastwind} \citep{Santolaya-Rey1997, Puls2005}, as well as information about the spectroscopic binary status of the complete sample of stars obtained from the available multi-epoch data~set \citep{Holgado2019}. In addition, we benefited from the theoretical predictions provided by two different single star evolutionary codes \citep{Brott2011, Ekstroem2012}, and the results of the binary population synthesis simulation presented by \cite{deMink2013}.

This paper is intended to broaden the scope of the works by \cite{Simon-Diaz2014} and \cite{Markova2014}, also improving and superseding the information presented in some important reference studies such as \cite{Conti1977}, \cite{ Penny1996}, and \cite{Howarth1997}. In addition, it can be considered as an extension of the investigation of the rotational velocities of a sample of presumably single and spectroscopic binary O-type stars in the 30~Doradus region of the Large Magellanic Cloud (LMC), which was performed by \cite{Ramirez-Agudelo2013, Ramirez-Agudelo2015} in the framework of the VLT-FLAMES Tarantula Survey \citep[VFTS,][]{Evans2011}.

The structure of this paper is as follows.
The sample, observations, and spectroscopic analysis tools used are briefly described in Sect.~\ref{sectionObsMeth}.
Results, in Sect.~\ref{sectionResults}, constitute the core of the paper, where we discuss the global properties of the sample and the interdependence between \vsini\ and evolution.
Section~\ref{sectionDiscuss} evaluates the empirical results, taking into account the trends that evolutionary models produce, with Sect.~\ref{ViniSect} focusing on the initial rotation distribution in O-type stars.
Concluding remarks and future prospects are included in Sect.~\ref{Sect_summary}.

\begin{table*}[!t]
        \caption{Main characteristics of several reference studies that investigate the rotational properties of high-mass stars in the Milky Way (MW), and the Large and Small Magellanic Clouds (LMC and SMC), covering the whole O-star domain.} 
        \label{ComparisonOther}
        \centering
        \begin{tabular}{cc@{\hskip 0.1in}c@{\hskip 0.1in}c@{\hskip 0.1in}c@{\hskip 0.1in}c}
                \hline \hline
        \noalign{\smallskip}
          Study & Galaxy   &  Sample size & \vsini/\vmacro  & Spectroscopic & Specific \\
                  &        & (O stars)    & separation      & parameters  & SB study \\
                \hline
        \noalign{\smallskip}        
        This work             & MW      & 285 & Yes & Yes & SB1       \\
                \hline
        \noalign{\smallskip}   
        \cite{Conti1977}       & MW     & 205 &  No & No           &  No  \\
        \cite{Penny1996}       & MW     & 177 &  No & No           &  No  \\
        \cite{Howarth1997}      & MW    & 225 &  No &  No & No            \\
        \cite{Simon-Diaz2014}       & MW   &  116 & Yes  & No & No           \\
        \cite{Markova2014}       & MW   &  39 & Yes  & Yes & No           \\
                \hline
        \noalign{\smallskip}   
        \cite{Penny2009} & SMC/LMC & 39/70 & No &  No & No  \\
        \cite{Ramirez-Agudelo2013, Ramirez-Agudelo2015} & LMC & 326 & Yes  &  No & SB1, SB2 \\                
                \hline
        \end{tabular}
\end{table*}

\section{Sample definition and methodology}\label{sectionObsMeth}

The final sample of targets considered for this work was drawn from an initial sample of 415 Galactic O-type stars, for which the IACOB and OWN projects have presently available $\sim$3900 high-resolution ($R$\,=\,25000\,--\,85000), multi-epoch spectroscopic observations. To avoid too much repetition, we refer the reader to \cite{Holgado2020} for a detailed description of how this sample was built. There, we also describe its main characteristics in terms of spectral type and luminosity class coverage, as well as its completeness when compared with the list of O-type stars quoted in version 4.1 of the Galactic O-star catalog \citep[GOSC,][]{MaizApellaniz2013}. 
 
From this initial sample, we excluded all those stars clearly identified as double-line spectroscopic binaries (SB2s, 113 objects) or Wolf-Rayet (4), as well as all the targets presenting features in their spectra that are typically associated with Oe (6), or magnetic (7) stars, for which a successful fitting with our {\sc fastwind} set of models was not attainable\footnote{Three stars from the final sample have measured magnetic fields present, but they present no caveats for the spectral fitting (HD~57682, HD~37742, and HD~54879)}. 
The main reason supporting this decision was this inherent limitation of our analysis strategy (see below) to provide reliable stellar parameters for these stars.

We also benefited from the multi-epoch character of the IACOB and OWN surveys to flag the stars in the remaining sample that could be clearly identified as single-line spectroscopic binaries (SB1s). To this aim, following the strategy presented in \cite{Holgado2018}, we obtained radial velocity (\vrad) estimates for all available spectra. Then, after computing the associated dispersion -- $\sigma$(\vrad) -- and the peak-to-peak amplitude -- $\Delta$\vrad\ -- of all measurements, we identified as clear SB1s the stars that fulfilled the criteria described in Appendix \ref{quan_var} \citep[see also][]{Holgado2019}.   

As a result, we ended up with a final sample of 285 stars, for which we were able to obtain their projected rotational velocities (\vsini) and the spectroscopic parameters. Among them, 55 stars were flagged as SB1s. The other 230 stars are referred to as likely single (LS), although some of them might still be undetected spectroscopic binaries and/or SB1 stars with low amplitude \vrad\ variations. In this regard, we remark that, by the time of finalizing this paper, we had only one epoch available for 54 out of the 230 stars in the LS subsample.

Details about the tools we used and the methodology we followed to perform the quantitative spectroscopic analysis of our working sample of 285 O-type stars can be found in \cite{Holgado2018, Holgado2020} and references therein. In brief, our analysis strategy, applied to the best signal-to-noise ratio (S/R) spectrum per star, can be summarized in two main steps. 

First, we obtained estimates for the two line-broadening parameters (\vsini\ and \vmacro) using the {\sc iacob-broad} tool \citep{Simon-Diaz2014} and following the guidelines presented in \cite{Holgado2018}. In particular, we based our \vsini\ determination on the \ioni{O}{iii}\,$\lambda$5592 line whenever possible (see, however, notes in Sect.~\ref{sectionResults}), and we assumed a radial-tangential definition of the macroturbulent broadening profile \citep[as suggested by][]{Simon-Diaz2014}. In addition, we benefited from the possibility offered by {\sc iacob-broad} to compare the \vsini\ values resulting from the Fourier Transform (FT) and goodness-of-fit (GOF) analysis, as an assessment of the reliability of the results regarding this quantity.

Next, we performed a HHe analysis\footnote{For some of the early O-type stars (in which \ioni{He}{i} lines were weak or absent), the effective temperature was alternatively constrained using several available \ioni{N}{iv-v} lines \citep[see][]{RiveroGonzalez2012, Holgado2020}.} using the semi-automated tool {\sc iacob-gbat} \citep{Simon-Diaz2011, Holgado2018}, which has incorporated a grid of synthetic spectra computed with the \fastwind\ \citep{Santolaya-Rey1997, Puls2005, RiveroGonzalez2012} stellar atmosphere code. As a result, we obtained estimates and associated uncertainties for the stellar effective temperature (\Teff), surface gravity (\grav), helium surface abundance (\helio), microturbulence (\micro), and wind-strength $Q$-parameter. The complete results of this analysis were presented in \cite{Holgado2019}, and a subset of the derived parameters can be found in \citet[][see also Table~\ref{tableValues}]{Holgado2020}.

In this paper, we mainly make use of three of the parameters determined spectroscopically: \vsini, \Teff, and \grav\footnote{Although for simplicity we use the term \grav\ along the paper, we actually refer to  the quantity \gravt, where $g_{\rm corr}$ is the surface gravity determined spectroscopically ($g_{\rm sp}$) corrected from centrifugal forces ($g_{\rm cent}$). In particular, $g_{\rm corr}$=$g_{\rm sp}$+$g_{\rm cent}$, and $g_{\rm cent}\approx(v\sin i)^2$/$R_{*}$ \citep[see notes in][]{Repolust2004}.}. Indeed, instead of \grav, we use the quantity $\mathcal{L} := T_{\rm eff}^4/g$, which allows us to locate the stars in the so-called spectroscopic Hertzsprung-Rusell diagram \citep[sHRD, ][]{Langer2014}. 

Similarly to the case of \cite{Holgado2020}, we preferred to use this diagram instead of the traditional Hertzsprung-Rusell diagram (HRD), given that 45 (out of the 285) stars in the sample present a value of renormalized unit weight error (RUWE)$>$1.4 in the $Gaia$ EDR3 release \citep{Lindegren2018,GaiaCollaboration2021}, and hence the reliability of available parallaxes for these stars still needs to be put under quarantine.
As discussed in \cite{Langer2014}, the sHRD can be considered as an analogous version of the HRD, but with the benefit of allowing the comparison of observations and evolutionary models regardless of distance and extinction constraints, since it is purely based on parameters determined spectroscopically.

\subsection*{Comparison with previous studies: Sample size and available empirical information}

As indicated in Sect.~\ref{sectionIntro}, one of the initial drivers of the study presented in this paper was to provide an updated and improved empirical overview of the spin-rate properties of Galactic O-type stars. Table~\ref{ComparisonOther} allows us to evaluate the main improvements that we have achieved with respect to several commonly referenced studies found in the literature. We cover four different aspects: (a) the size of the investigated sample of stars; (b) whether the study uses \vsini\ estimates decontaminated from the effect of the so-called macroturbulent broadening; (c) the availability of \Teff\ and \grav\ estimates; and (d) whether the study makes any specific study comprising the spectroscopic binaries that are present in the sample under study.

As can be concluded from examination of Table~\ref{ComparisonOther}, the empirical information provided in (and discussed throughout) this paper implies a clear improvement -- in at least two of the aspects mentioned above -- with respect to all quoted studies about rotational velocities in Galactic O-type stars. More specifically, our working sample of stars is not only $\sim$20\,--\,40\% larger\footnote{Actually, $\sim$40\,--\,50\% larger, if we also take into account the discarded SB2 and peculiar stars (see Sect.~\ref{sectionObsMeth}).} than those considered by \cite{Conti1977}, \cite{Penny1996}, or \cite{Howarth1997}, but we have also been able to: (a) supersede their derived \vsini\ with new (more robust) estimates of this quantity; (b) better identify spectroscopic binaries; and (c) routiney incorporate information about \Teff\ and \grav\ in all stars in the sample that are not identified as SB2s.

Regarding the more recent works by \cite{Simon-Diaz2014} and \cite{Markova2014}, our paper can be seen as a natural extension of these two works.  Here, we investigate a considerably larger sample of stars, with a better coverage in terms of spectral type and luminosity class\footnote{Also eliminating some specific limitations (in terms of mass coverage) affecting the sample of \cite{Markova2014}.}, and with freshly incorporated information about the spectroscopic binary status for a large fraction of them.

\section{Results}\label{sectionResults}

\subsection{General overview}\label{GenOv}

Table~\ref{tableValues} summarizes the compiled empirical information of interest for our working sample of 285 Galactic O-type stars\footnote{See also some notes about the 13 peculiar stars (Oe and magnetic) found in the initial sample in Appendix~\ref{Peculiar}.}. We repeat the same data provided in \cite{Holgado2020} regarding spectral classification (as quoted in GOSC), spectroscopic  parameters -- \Teff, \grav, \Lsp\ --, and whether the star was identified as line profile variable (LPV)  or SB1, using the set of available multi-epoch spectra. In addition, we complement this information with the new listed derived line-broadening parameters \vsini\ and the \vmacro\ used for this work\footnote{Although the line-broadening parameters of the full sample of O-type stars considered here were already calculated and taken into account for the results presented in  \citet[][]{Holgado2020}, this is the first time they are quoted in a table.}, also indicating the diagnostic line used in the {\sc iacob-broad} analysis. Similarly to the table presented in \cite{Holgado2020}, stars in Table~\ref{tableValues} are grouped by luminosity class (LC) and ordered by spectral type (SpT).

\begin{figure}[!t]
\includegraphics[width=0.5\textwidth]{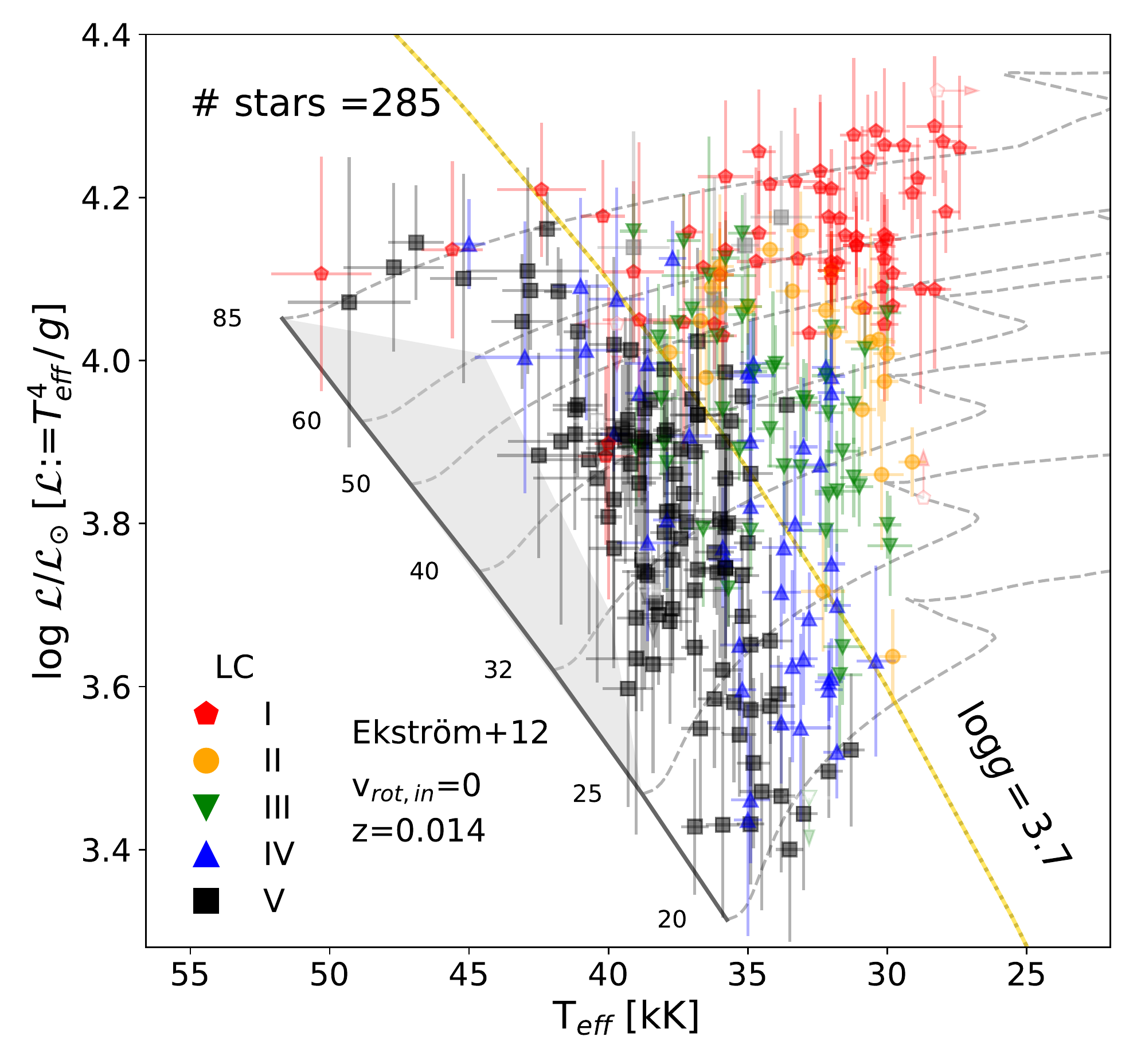}
\caption{Spectroscopic HR diagram indicating the location of the 285 Galactic O-type stars investigated in this work. Different colors and symbols are used to highlight luminosity classes. Individual uncertainties are included as error bars. Evolutionary tracks and position of the ZAMS from the non-rotating, solar-metallicity models by \cite{Ekstroem2012}, plus the \grav=3.7 dex line of constant gravity, 
are also included for reference purposes. The gray shadowed area highlights a region of the diagram which is void of stars \cite[see main text and][for further explanations]{Holgado2020}.
}
\label{Gen_Sample_HR}
\end{figure}

\begin{figure}[h]
\includegraphics[width=0.5\textwidth]{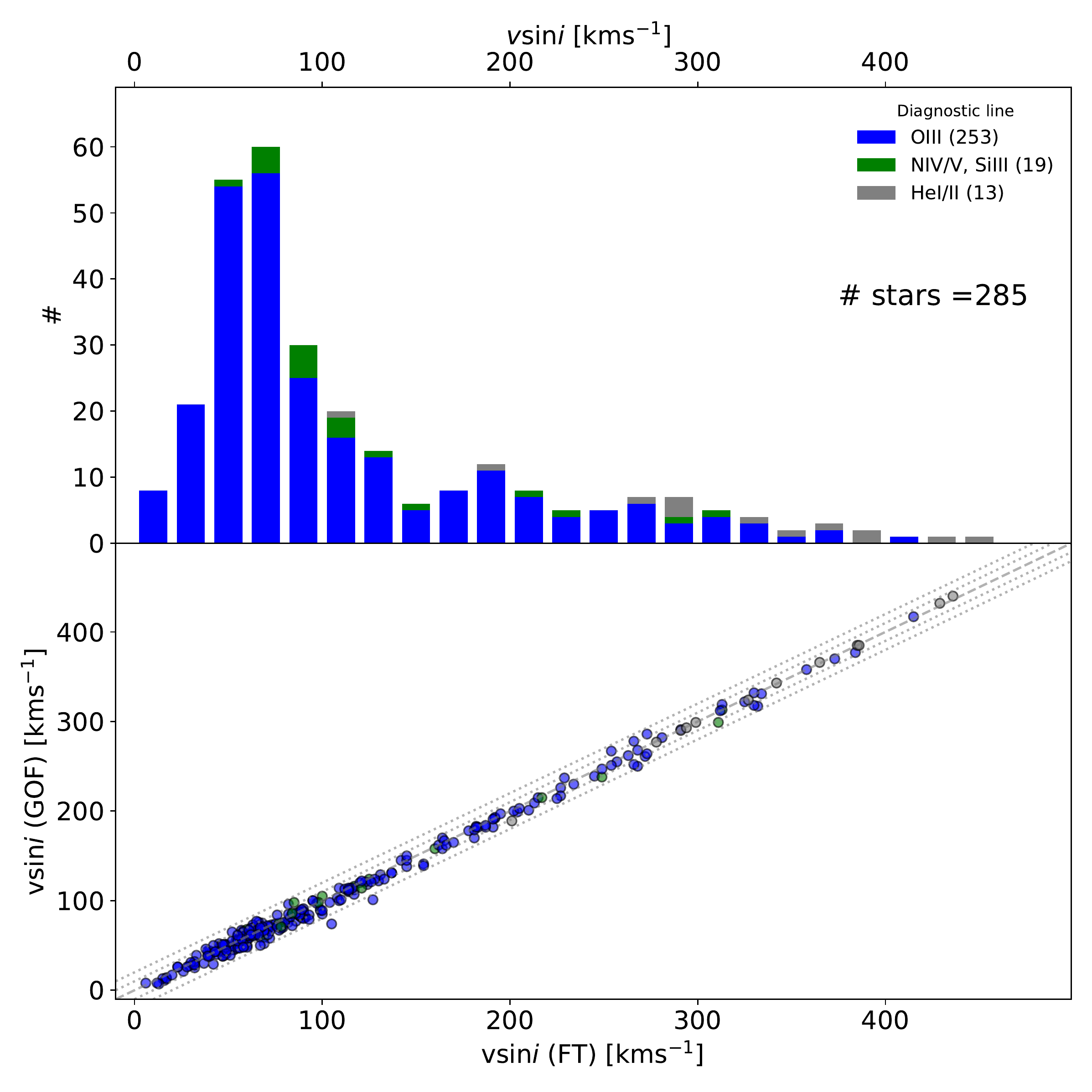}
\caption{Rotational properties of the sample. \textit{(Top)} Histogram of \vsini\ values resulting from the {\sc iacob-broad} analysis of the 285 O-type stars in our working sample, highlighting the diagnostic line considered for the analysis. \textit{(Bottom)} Comparison of the \vsini\ estimates resulting from the FT and GOF analysis strategies. The diagonal dotted lines indicate differences between both quantities of $\pm$10 and $\pm$20~\kms.}
\label{GOF_FT}
\end{figure}

\begin{figure}[h]
\includegraphics[width=0.5\textwidth]{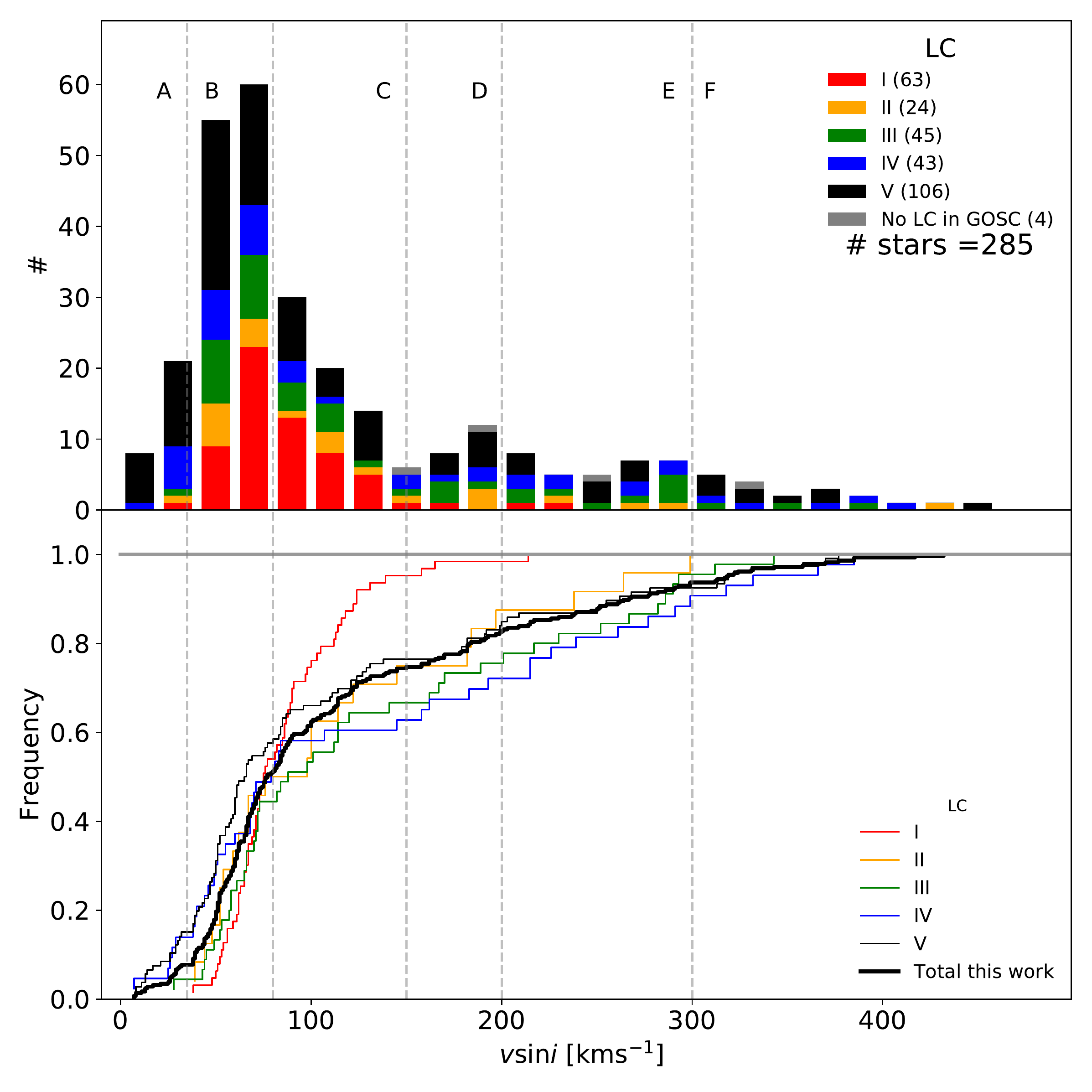}
\caption{Histogram (\textit{top}) and cumulative distribution (\textit{bottom}) of \vsini\ estimates for the 285 O-type stars investigated in this work, separating results for the various luminosity class subgroups (same color code as in Fig.~\ref{Gen_Sample_HR}).
Ranges in \vsini\ labeled with letters ``A'' to ``F'' are used in Fig.~\ref{cuadrvsini} and explained in Sect.~\ref{vsinisHR}.
}
\label{Gen_Sample_Hist}
\end{figure}

Figures~\ref{Gen_Sample_HR} to~\ref{Gen_Sample_Hist}  provide a first general overview of the location of the sample of investigated stars in the sHRD and the associated \vsini\ distribution. 
We refer the reader to \cite{Holgado2020} for a thorough discussion about the general distribution of the sample in the sHRD, including possible explanations for the scarcity of stars close to the zero-age main sequence (ZAMS) in the mass range $\sim$30\,--\,80~\msol\ (marked in gray in Fig.~\ref{Gen_Sample_HR}).
Also, while we do not consider it necessary to provide additional information about the reached level of accuracy and reliability of the measured effective temperatures and surface gravities \cite[since this is already properly discussed in][]{Holgado2018}, we indicate below some information of interest regarding our \vsini\ measurements.
In addition, a discussion of some potential caveats and limitations that could still affect the reliability of the estimates of this quantity in the low- \vsini\ regime can be found in Sect.~\ref{lowvalues_intext}.

As indicated in Sect.~\ref{sectionObsMeth}, for the sake of homogeneity, we tried to mainly base our \vsini\ determinations on the \ioni{O}{iii}\,$\lambda$5592 line. This was possible for almost 90\% of the stars in the sample (see top panel in Fig.~\ref{GOF_FT}). There were, however, a few late and early O-type stars for which the \ioni{O}{iii} line was weak and we had to rely on alternative lines. In most of those cases, we were able to still use a metal line (either \ioni{Si}{iii}\,$\lambda$4552, \ioni{N}{iv}\,$\lambda$6380, or \ioni{N}{v}\,$\lambda$4603/20\footnote{Being aware that the \ioni{N}{v}\,$\lambda$4603/20 lines can be affected by  wind, they were only used if the absence of a strong wind was confirmed.}), allowing us to reach 95\% of the stars. But, for some very fast rotators (37 stars, 5\% of the total sample) all metal lines appeared too diluted to provide reliable results, and we had to rely on \ioni{He}{i} lines. Although the use of a \ioni{He}{i} line may present caveats in the case of slow rotators \citep[especially in analyses based on medium and low-resolution spectra, or when an important nebular contamination affects the \ioni{He}{i} lines, see, e.g.,][]{Ramirez-Agudelo2013}, we note that almost all the stars for which we needed to use a \ioni{He}{i} line have a \vsini\ value above 200~\kms. As a result, in these stars the effect of Stark broadening in the determination of \vsini\ can be considered as negligible. Also, none of the analyzed \ioni{He}{i} lines presented any nebular contamination.


The bottom panel of Fig.~\ref{GOF_FT} shows the good agreement found between the two analysis strategies provided by {\sc iacob-broad}, with the comparison of \vsini(FT) and \vsini(GOF) estimates . Differences are smaller than 10~\kms\ in $\sim$85\% of the sample and, except for a few stars, are not larger than 15~\kms\ for the rest. This result can be used as an assessment of the reliability of the derived \vsini\ values, since they were obtained using two complementary analysis strategies (see, however, Sect.~\ref{lowvalues_intext}).

In view of this result, we decided to use \vsini(GOF) estimates and to establish 5\,--\,15~\kms\ as the typical (formal) uncertainty associated with the measurement of this quantity from our high-quality spectroscopic data~set. This is also the reason why we use a bin size of 20~\kms\ in the \vsini\ histograms presented hereafter throughout the paper. See Appendix~\ref{AppBines} for a review on the significance of this selection. 


\subsection{The empirical \vsini\ distribution of Galactic O-type stars}

\subsubsection{Properties of the global \vsini\ distribution}\label{bimodalSect}

Figure~\ref{Gen_Sample_Hist} depicts the information about the \vsini\ distribution of our sample of 285 Galactic O-type stars in the form of a histogram (top) and a cumulative distribution function (bottom). In addition to the global distribution, we also provide the resulting distributions separated by luminosity class groups. This helps us to establish a first link between our results and those presented in previous investigations about the rotational distribution of O-type stars.

Inspection of these two panels allows us to point out two main features already highlighted in the seminal work by \cite{Conti1977}, and later on reproduced by other similar studies, such as those by \cite{Penny1996} and \cite{Howarth1997} for the case of Galactic stars, or \cite{Penny2009} and \cite{Ramirez-Agudelo2013, Ramirez-Agudelo2015} for the case of stars in the Magellanic Clouds.
On the one hand, the global distribution shows a clear bimodal character, with a slow velocity peak comprising most of the stars (in our case, $\sim$75\% of the total sample, if we consider that this part of the distribution ends in \vsini$\sim$150~\kms), and a tail of fast rotators, reaching \vsini\ values
up to $\sim$450~\kms. On the other hand, the ranges in \vsini\ covered by the various luminosity class groups are different, with the two most extreme cases found between the dwarfs (LC~V) and the supergiants (LC~I). While the former are distributed over the whole range, the latter are mostly concentrated within the low-velocity peak feature of the distribution.
This feature is well-documented also for other spectral classes \citep[e.g.,][for A- and B-type stars, respectively]{Royer2002,Dufton2013}.

Several studies have proposed that for less massive stars, the origin of this bimodality may lie in their ability to maintain, or not, the presence of a disk of material around themselves, producing an interaction between the stars and their accretion disks.\ This interaction is able to remove angular momentum and prevent the spin up, and is normally simplified as the concept of “disk locking” \citep[see][and references therein]{Bastian2020}.
In the cases of stars born with masses in the range covered by the O-type stars, the bimodal character of the \vsini\ distribution (mainly present in the dwarf sample, and initially speculated to reflect the initial spin distribution, see Fig.~\ref{Gen_Sample_HR}) could not be convincingly explained in the early works by \cite{Conti1977}. However, more recent studies have proposed that this feature is a natural consequence of the effect of binary interaction during massive star evolution \citep[][see also Sect.~\ref{BinInteract}]{deMink2013}.
Therefore, taking into account the large percentage of O-type stars that are expected to evolve as part of a binary system \citep{Sana2012}, the incorporation of information about the spectroscopic binary status of individual targets in any global empirical study of the rotational properties of massive stars has become of crucial interest.

\subsubsection{Likely single stars versus single-line spectroscopic binaries}\label{LSySB1}

To the best of our knowledge, there is only one comprehensive study that performs a detailed comparison of the rotational properties of a statistically meaningful sample of single and spectroscopic binary stars in the O-star domain. We refer to the work by \cite{Ramirez-Agudelo2015} in the 30 Doradus region of the LMC.
No similar study has been carried out yet for Galactic O-type stars. 

Thanks to the multi-epoch character of the IACOB and OWN surveys, we can take the first steps in this direction.
As mentioned in Sect.~\ref{sectionObsMeth}, the compiled spectroscopic data set allowed us to identify about 170 spectroscopic binaries (including 113 SB2 systems and 55 clearly detected SB1 stars) within our initial sample of 415 Galactic O-type stars. The rest of the stars in the sample (230, excluding 17 peculiar stars) have been flagged as likely single. While the SB2 systems will not be further considered for this work (see Sect.~\ref{sectionObsMeth}), throughout this paper we also investigate similarities and differences that can be found between our LS and SB1 subsamples.

\begin{figure}[!t]
\includegraphics[width=0.5\textwidth]{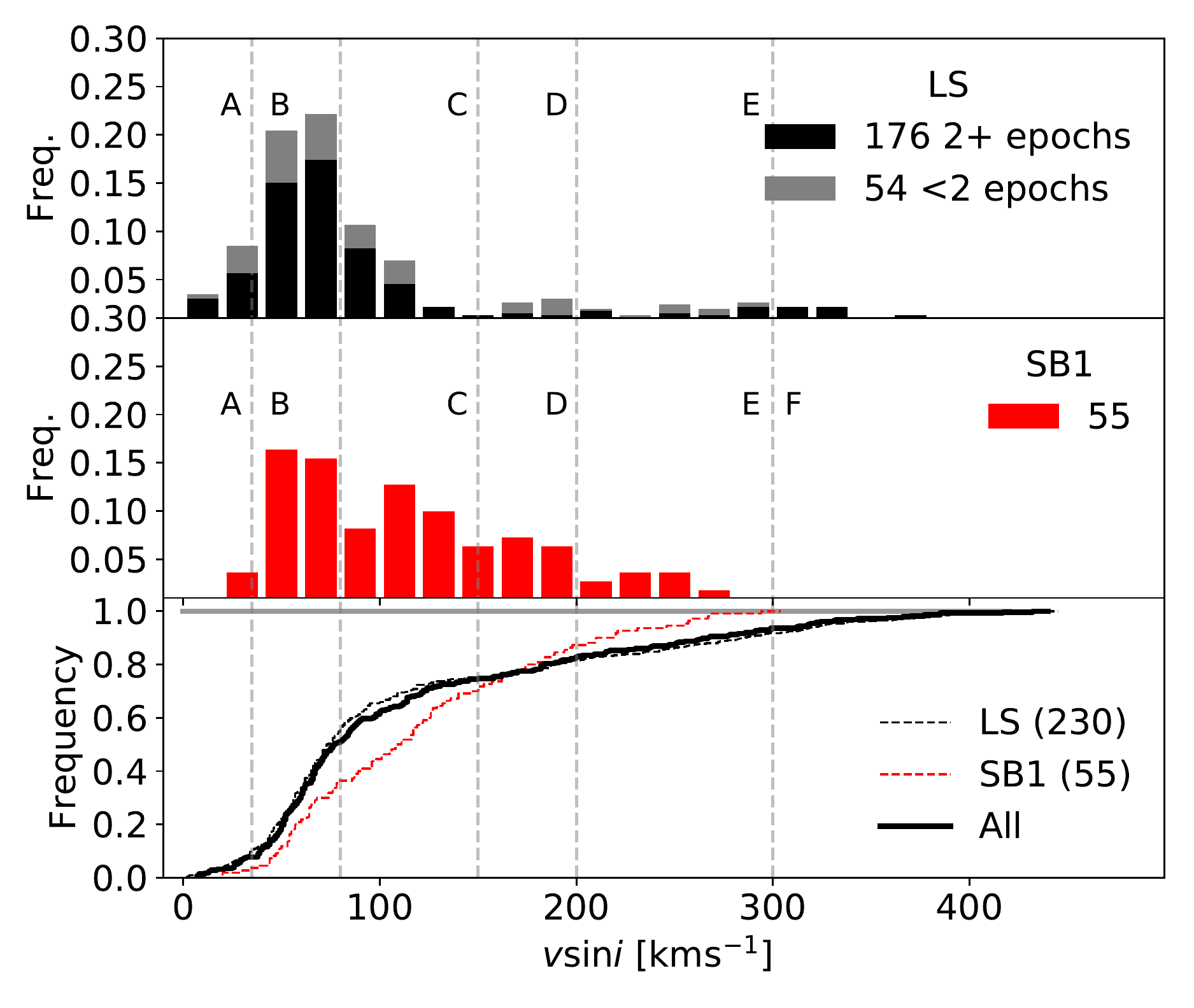}
\caption{ 
Comparison of the \vsini\ distributions of the likely single (LS) and clearly detected SB1 subsamples, presented as histograms \textit{(top and middle panels, respectively)} and cumulative distributions \textit{(bottom panel)}.
Stars in the LS sample, for which we only have one spectrum, are highlighted in gray.
Ranges in \vsini\ labeled with letters ``A'' to ``F'' are used in Fig.~\ref{cuadrvsini} and explained in Sect.~\ref{vsinisHR}.
}
\label{RotCompare_CSB1_VH}
\end{figure}

In view of this, Fig.~\ref{RotCompare_CSB1_VH} depicts the \vsini\ distributions resulting from these two subsets of stars separately.
As illustrated by the lower panel of Fig.~\ref{RotCompare_CSB1_VH}, the distribution of likely single stars is very similar to the combined one, meaning that they clearly dominate the global distribution (as expected from the relative percentage of stars in each subsample).
In contrast, the comparison of the LS and SB1 samples highlights clear differences. 
In fact, a Kuiper's test (Kuiper 1960) indicates that there is a very low probability (below 1\%) that the two observed distributions are randomly drawn from the same parent population.

More specifically, the bimodal character of the global \vsini\ distribution is not so clearly defined in the case of the SB1 sample. Indeed, one could assume that there is only one component that, compared to the main (low \vsini) component of the LS star distribution, is wider (by a factor of $\sim$2) and it is centered at a somewhat larger \vsini\ (120 vs. 65~\kms).
In addition, there is a lack of SB1 stars with projected rotational velocities larger than 300~\kms.

Interestingly, similar results were obtained by \cite{Ramirez-Agudelo2015} when they compared the observed distributions of the presumed single stars and the primary components of those spectroscopic binaries comprising the O-type star population of the 30 Doradus region of the LMC. Following \cite{deMink2013}, these authors propose that the observed differences are most likely produced by a combination of effects taking place along the various stages of early evolution of high-mass binaries (such as tides, mass and/or angular momentum transfer, and magnetic braking; see also additional notes in Sect.~\ref{BinInteract}).
For a more detailed comparison of our resulting \vsini\ distribution with the one obtained by \citeauthor{Ramirez-Agudelo2013}, we refer the reader to Appendix~\ref{RAGUAppend}.


\begin{figure*}[!t]
\includegraphics[scale=0.38]{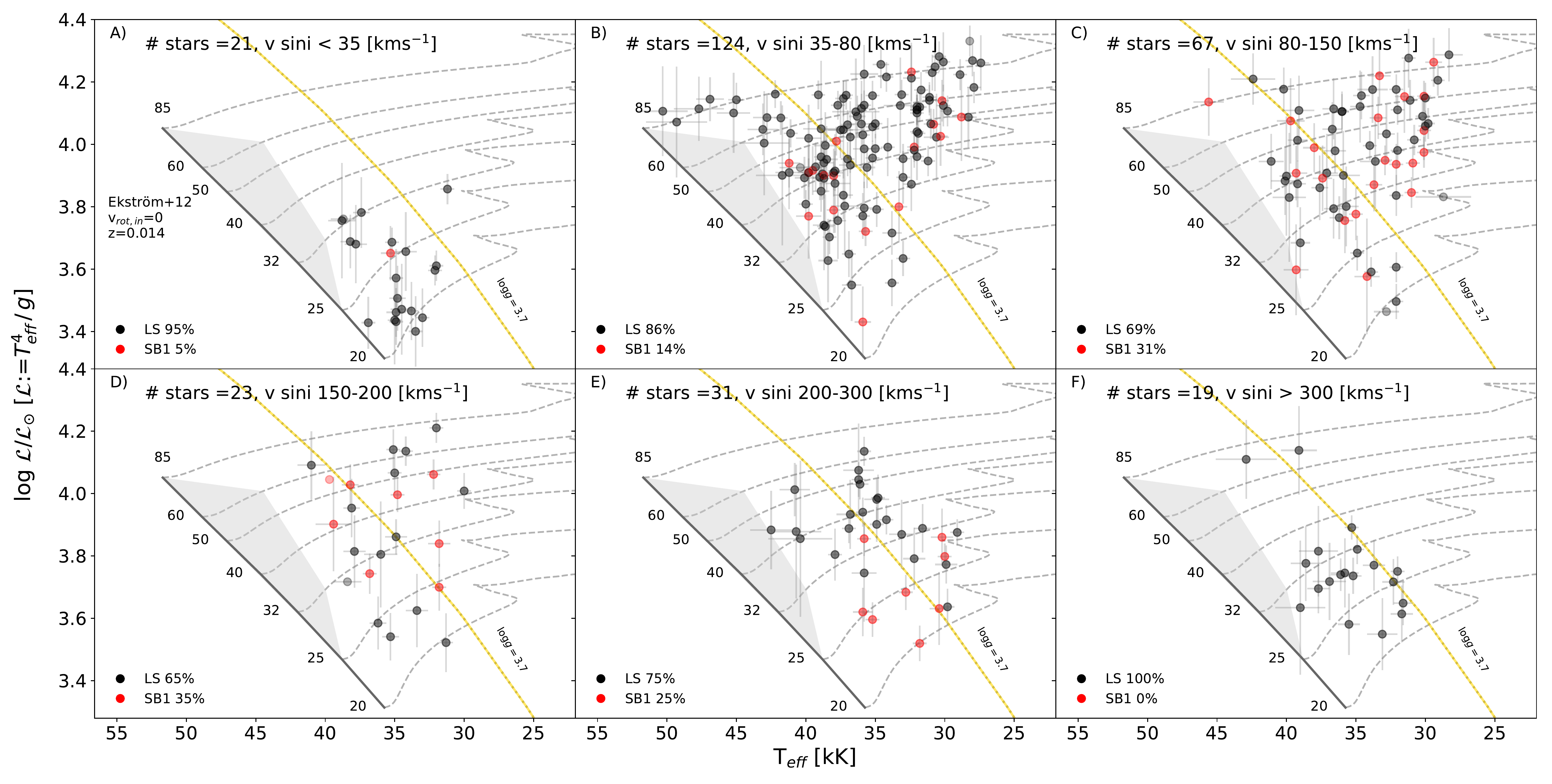}
\caption{Location in the sHRD of our working sample of stars as a function of the various \vsini\ ranges indicated in Figs.~\ref{Gen_Sample_Hist} and \ref{RotCompare_CSB1_VH} (see also Sect.~\ref{vsinisHR}). The LS and SB1 subsamples are highlighted in black and red, respectively, and their relative percentage in each panel is included for reference. Further information about other features presented in this figure can be found in the caption of Fig.~\ref{Gen_Sample_HR}.
}
\label{cuadrvsini}
\end{figure*}

\subsubsection{The low- \vsini\ regime}\label{lowvalues_intext}

In previous sections, we have indicated that our revised \vsini\ estimates can be considered as more robust and reliable than those provided by previous works \citep[e.g.,][]{Conti1977, Howarth1997, Penny1996}. This is mainly a result of the combined use of a high-quality spectroscopic data set (in terms of spectral resolution and S/R) and a line-broadening analysis technique, allowing us to disentangle the effect of the rotational and the so-called macroturbulent broadening components. However, following \cite{Sundqvist2013}, \cite{Simon-Diaz2014}, and \cite{Markova2014}, we are aware that our analysis methodology might still present some limitations in the low- \vsini\ regime.

This potential failing of the considered line-broadening analysis technique clearly stands out from inspection of the \vsini\ distribution presented in Fig.~\ref{Gen_Sample_Hist}: there is a notorious deficiency of O stars with spectroscopically inferred \vsini\ below $\sim$40-50~\kms\ (particularly in the case of stars with LCs I, II and III). Taking into account that our instrumental set-up imposes a lower limit in the measured \vsini\ of 5\,--\,10~\kms, plus the statistical effect of the inclination angle ($i$), one would expect a larger number of stars in the low- \vsini\ region of the distribution.

As extensively discussed in \cite{Simon-Diaz2014} and references therein, this long-standing problem -- which was already present (to a larger extent) in the works by \cite{Conti1977, Howarth1997}, or \cite{Penny1996} -- has been only partially alleviated by accounting for the effect of macroturbulence on the determination of projected rotational velocities. Other sources of line-broadening seem to still be hindering the suitability of available methods to reach actual \vsini\ measurements below $\sim$40\,-\,50~\kms\ in some specific situations.

While there is not yet a definitive answer as to what is the main culprit of this methodological limitation, some empirical hints seem to point toward so-called microturbulent broadening. As pointed out by \cite{Gray1973}, classical atmosphere microturbulence also produces zeroes in the FT at (high) frequencies associated with low values of \vsini\ that may be falsely identified as the zeroes associated with the rotational broadening below a certain \vsini\ limit. In this context, \cite{Simon-Diaz2014} showed that microturbulence values in the range 5\,--\,25~\kms\ can lead to Fourier space minima at frequencies corresponding to \vsini\,$\sim$\,10\,--\,40~\kms. The good correlation between the Fourier and GOF methods shown in Fig.~\ref{GOF_FT} indicates that the latter is probably also affected by this unaccounted broadening.

Some additional notes about this line of research can be found in Appendix~\ref{lowvalues}, where we also summarize the main results obtained in the context of this paper, which could help to shed light on this problem. Hereafter, throughout the paper, we take this warning into account for any interpretation of results. In particular, we consider (as a first order approach) that any \vsini\ estimate between $\sim$40\,--\,50~\kms, could actually correspond to an upper limit of this quantity, especially in the upper part of the sHRD, or in the case of stars for which a high value of \vmacro\ has been obtained \cite[following the results presented in Fig.~4 of][]{Simon-Diaz2017}.

\subsection{The \vsini\ distribution of Galactic O-type stars in the sHRD}\label{vsinisHR}

In most previous studies about the rotational properties of O-type stars, spectral types and luminosity classes were used as a proxy of mass and evolutionary status, since direct information about the stellar parameters of the investigated samples was not available.
Having access to \Teff\  and \grav\ estimates allows us to take one step further in the description and interpretation of results \citep[see, e.g.,][]{Markova2014}.

To serve as a first example, Fig.~\ref{cuadrvsini} shows (in six separated panels) how stars covering different ranges of the global \vsini\ distribution populate the sHRD. In each panel, we differentiate those stars identified as SB1.
The considered ranges (indicated in Figs.~\ref{Gen_Sample_Hist} and \ref{RotCompare_CSB1_VH} with letters "A" to "F"), were specifically selected to separate several regions of interest within the two main components of the global (bimodal) \vsini\ distribution. In particular, regions "A", "B", and "C" (depicted in the top panels of Fig.~\ref{cuadrvsini}) cover the low-velocity peak of the distribution, ranging up to 150~\kms\ and including $\sim$75~\% of the stars in the working sample. Regions "D", "E", and "F" (bottom panels) correspond to the tail of fast rotators. Although the assumed boundaries for the various regions within each of the two main components could be initially seen as somewhat arbitrary, they were actually chosen to illustrate several features of interest that emerge when considering how the stars in each \vsini\ range are distributed in the sHRD.

The most evident cases correspond to panels A and F, which comprise the two extreme regions of the \vsini\ distribution. On the one hand, the upper \vsini\ boundary of region A has been specifically selected to show the lack of stars with measured \vsini\ below $\sim$35~\kms\ above the 32~\msol\ track. We note that the value of 35~\kms\ is used here only for that particular reason, as stars with low \vsini\ are described throughout the paper as those with \vsini$<$40-50~\kms. On the other hand, we find it interesting to remark how the (19) stars with \vsini~$>$~300~\kms\ (panel F) are mostly concentrated in the lower-mass region of the O-star domain (except for two targets\footnote{HD~14442 (O5n(f)p) and HD~229232 (O4\,V), both identified as LPV in Table~\ref{tableValues} \citep[see also][for the case of HD~229232, where the previously proposed SB1 status was challenged and also modified to single pulsating variable]{Trigueros2021}.}, all of them are located below the 40~\msol\ evolutionary track). Furthermore, no SB1 is detected among this latter sample, even though we have a minimum of five epochs for most of them (see Table~\ref{tableValues_over300}). 

A more careful inspection of the various panels in Fig.~\ref{cuadrvsini} allows us to highlight another group of stars which deserves further attention. We refer here to the bunch of (seven) stars closer to the upper part of the ZAMS (i.e., those with \Teff$>$45000~K and log($\mathcal{L}$/$\mathcal{L}_{\odot}$)$\sim$4.1~dex, see also Table~\ref{tableValues_over45K}), which are almost exclusively found in panel B (except for the only SB1 star in this group, located in panel C). This result was already pointed out by \cite{Markova2014}, who found suggestive evidence that stars with initial masses $\gtrsim$50~\msol\ rotate with velocities  not exceeding 26\% of their critical rotational velocities when they are close to the ZAMS. 
We note, however, that this sample of seven stars is still not statistically large enough to provide a robust confirmation of the lack of fast-rotating stars in this region of the sHRD, especially taking into account that we are actually measuring \vsini\ and not the surface equatorial velocity.

Also, it can be noticed that the upper right region of the sHRD (corresponding to the more evolved O-Supergiants) is void of stars in panel E, and only a few stars in this region are found in panel D. This result is coherent with Fig.~\ref{Gen_Sample_Hist}, where it is shown that the relative percentage of O stars with luminosity classes I and II drops drastically above \vsini$\sim$150~\kms, and becomes zero above $\sim$250~\kms (see also the specific location of the sample of LC I and II stars in the sHRD in Fig.~\ref{Gen_Sample_HR}).

Last, but not least, Fig.~\ref{cuadrvsini} offers us an interesting overview of the relative percentage (and distribution) of stars labeled as SB1 in each of the considered \vsini\ ranges. In particular, the relative percentage of SB1 stars reaches its maximum ($\sim$30\,--\,35\%) in panels C and D (\vsini\,=\,80\,--\,200~\kms), and decreases significantly below and above these limits. This is just a consequence of the different shape of the \vsini\ distributions of the LS and SB1 samples (see Fig.~\ref{RotCompare_CSB1_VH}).

The unexpected distribution of stars in panel A obviously reminds us that sometimes our presently available methods are still not suitable to provide reliable \vsini\ estimates in the low- \vsini\ regime (see notes in Sect.~\ref{lowvalues_intext} and Appendix~\ref{lowvalues}). Other than that, the results summarized above are expected to provide important empirical constraints for any attempt to investigate, from a theoretical point of view, the impact of high-mass star formation and (single and binary) evolution on the spin-rate properties of main sequence O-type stars (see further notes in Sect.~\ref{sectionDiscuss}).

%
\begin{figure}[!h]
\includegraphics[width=0.5\textwidth]{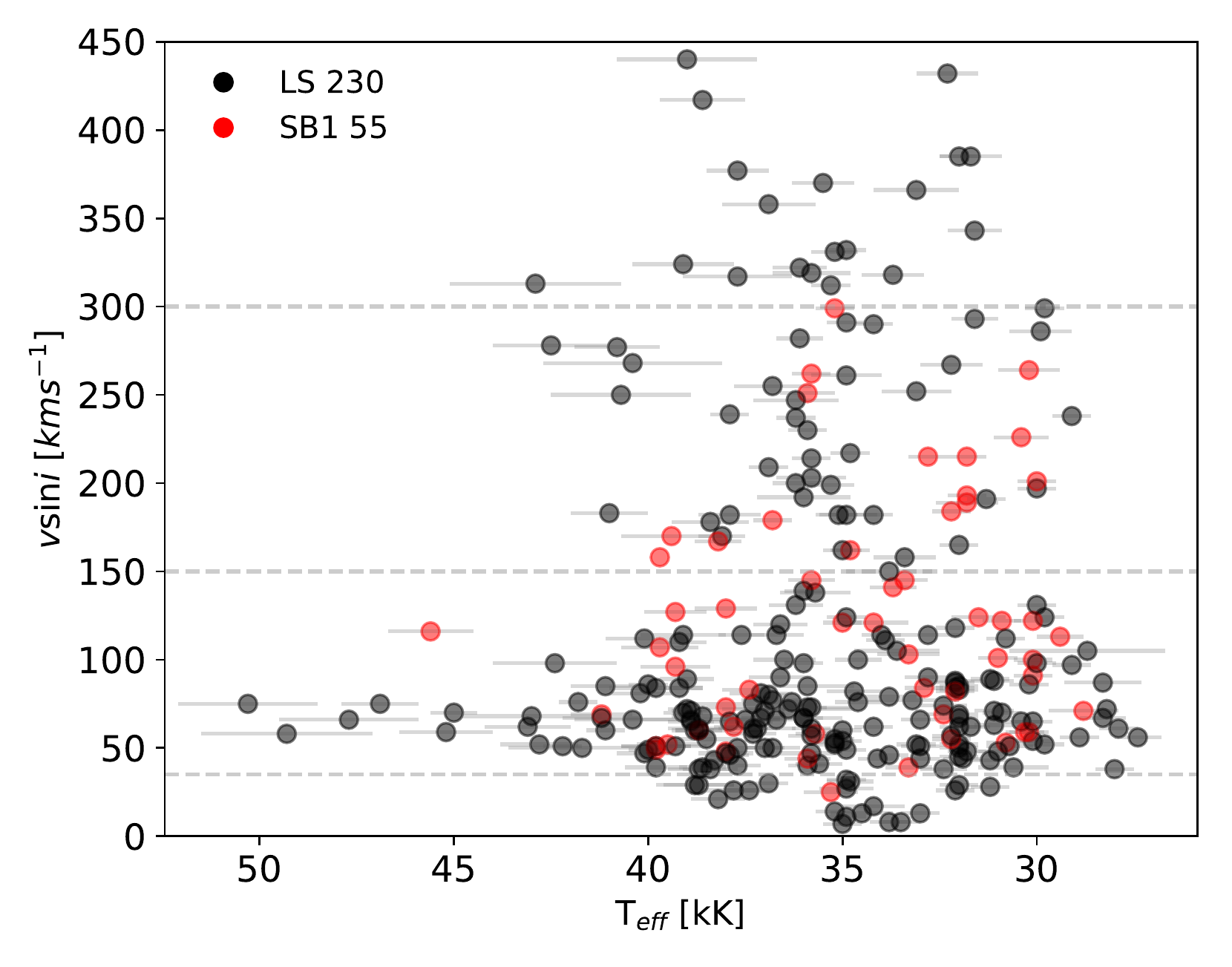}
\caption{\vsini\,--\,\Teff\ diagram including our working sample of stars. The LS and SB1 subsamples are highlighted in black and red, respectively.
The \vsini$\,=\,$35, 150, and 300~\kms\ values are highlighted for reference purposes (see Sect.~\ref{vsini_evol_Teff}).}
\label{Teff_vs_vsini}
\end{figure}

\subsection{\vsini\ dependence with \Teff\ and \Lsp}\label{vsini_evol_Teff}

\begin{figure}[!h]
\includegraphics[width=0.5\textwidth]{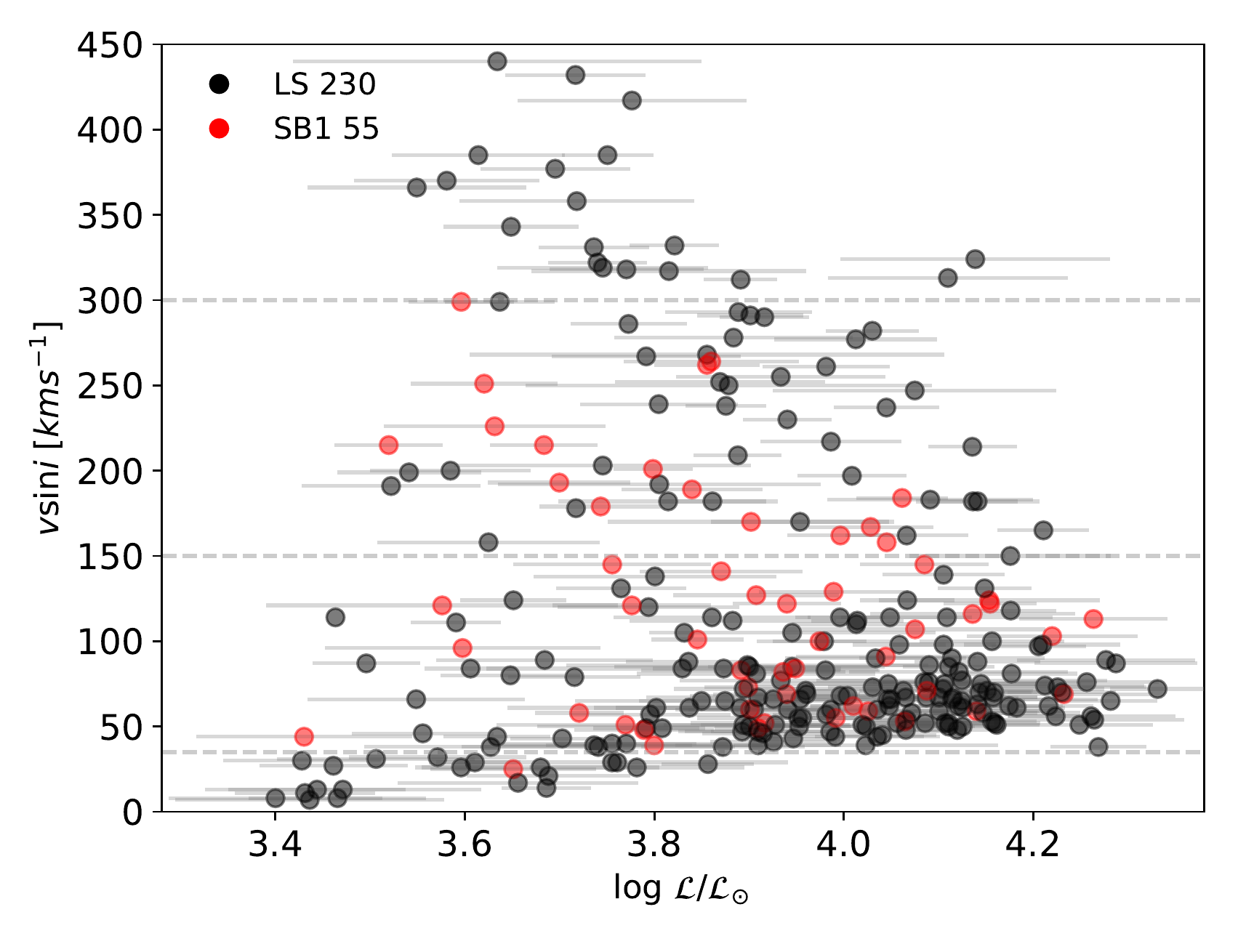}
\caption{\vsini\,--\,\Lsp\ diagram including our working sample of stars. The LS and SB1 subsamples are highlighted in black and red, respetively.
The \vsini$\,=\,$35, 150, and 300~\kms\ values are highlighted for reference purposes (see Sect.~\ref{vsini_evol_Teff}).
}
\label{logLsp_vs_Vsini}
\end{figure}

Figures \ref{Teff_vs_vsini} and \ref{logLsp_vs_Vsini} offer a complementary view of the information presented in Fig.~\ref{cuadrvsini} and described in the previous section.
In particular, the type of diagram presented in Fig.~\ref{Teff_vs_vsini} (either using spectral types or directly using \Teff\ in the x-axis\footnote{Some works also use \grav\ as a proxy of evolution in this diagram instead of \Teff\ \citep[see, e.g.,][]{Brott2011, Martins2013, Martinet2021}.}) has traditionally been used as a first approach to investigate the behavior of the (projected) rotational velocity of stars in the high-mass domain as a function of their evolution \cite[see, e.g.,][] {Conti1977, Fraser2010, Vink2010, Markova2014, Simon-Diaz2014, Keszthelyi2017}.

Thanks to the increased statistics, combined with the direct spectroscopic determination of the effective temperatures, our sample offers a more complete and robust overview (compared to previous attempts) of how Galactic O-type stars are distributed in this diagram. For example, in the case of \cite{Vink2010}, which used the long sample of O stars investigated by \citet[][see also Table~\ref{ComparisonOther}]{Howarth1997}, effective temperatures were obtained by converting spectral types, using SpT\,--\,\Teff\ calibrations. In addition, the considered \vsini\ estimates were still affected from so-called macroturbulent broadening. More recently, \cite{Markova2014} used a similar approach to the one utilized here to get estimates of \vsini\ and \Teff; however, their sample only included 39 O-type stars (see Table~\ref{ComparisonOther}) and, in their own words, was importantly affected by some selection effects.

The \vsini\,-\,log\,$\mathcal{L}$ diagram (Fig.~\ref{logLsp_vs_Vsini}) has not been so commonly used in previous works, but we also include it here because of its utility for the correct interpretation of some specific features shaping the distribution of stars in the \vsini\,-\,\Teff\ diagram. In addition, it also provides interesting clues to aid in the endeavor to identify the origin of the still remaining methodological limitations hindering a reliable determination of \vsini\ in the low-velocity regime.

%

Indeed, Fig.~\ref{logLsp_vs_Vsini} clearly shows that there exists a strong correlation between the lower limit in \vsini\ imposed by our analysis methodology and the quantity log~$\mathcal{L}$. As discussed in more detail in Appendix~\ref{lowvalues} (see also Sect.~\ref{lowvalues_intext}), this result indirectly supports the hypothesis that in low-\vsini\ stars, we might be misidentifying zeroes in the FT that are produced by the line-broadening effect of microturbulence, as if they were associated with rotational broadening.

Once we are aware of this correlation, we can better interpret the curious shape defining the lower boundary of the distribution of points in the \vsini\,-\,\Teff\ diagram (Fig.~\ref{Teff_vs_vsini}).
As seen in Fig.~\ref{Gen_Sample_HR}, the range in \Teff\ covered by stars with a given value of log~$\mathcal{L}$ becomes smaller as we move down in the sHRD. If we combine this characteristic distribution of our sample of O-type stars in the sHRD with the abovementioned correlation, we can understand why the lower \vsini\ limit in the \vsini\,-\,\Teff\ diagram reaches a minimum at \Teff$\sim$35\,000~K, and tends to increase toward both higher and lower effective temperatures. This is also the reason why stars in Panel A of Fig.~\ref{cuadrvsini} only populate the region below \Lsp$\sim$3.8~dex.

Beyond that, the most noticeable feature that stands out from inspection of Fig.~\ref{Teff_vs_vsini} is the concentration of stars with \Teff\,$\gtrsim$\,45\,000~K around the \vsini\ range between $\sim$50 and $\sim$75~\kms\ (except for one SB1 star which has a somewhat larger \vsini). Interestingly, when combined with the information presented in Fig.~\ref{cuadrvsini} (see panels B and C), it becomes clear that this appendix in the \vsini\,--\,\Teff\ distribution corresponds to the bunch of stars located closer to the ZAMS around the 85~\msol\ track. 
A similar distribution appears in \cite{Berlanas2020} for a sample of Cygnus OB2 OB stars.
This lack of fast rotators at the hot end of the distribution -- assuming that it is not an inclination effect -- together with the clear decrease in the upper \vsini\ boundary of the distribution of points in the \vsini\,-\,log\,$\mathcal{L}$ diagram (Fig.~\ref{logLsp_vs_Vsini}), could be interpreted as empirical evidence of the increasingly important braking effect produced by stellar winds, as more luminous stars are considered. However, as we will show in the next sections, this is not necessarily the correct interpretation, especially if we take into account, following \cite{deMink2013}, that a high fraction of the stars in the sample could actually be post-interacting binaries (even if they are currently detected as likely single stars). The work by \cite{Berlanas2020} also proposes a similar direction of investigation, with alternative evolutionary channels  proposed to explain these objects.


\begin{figure*}
\includegraphics[width=1.0\textwidth]{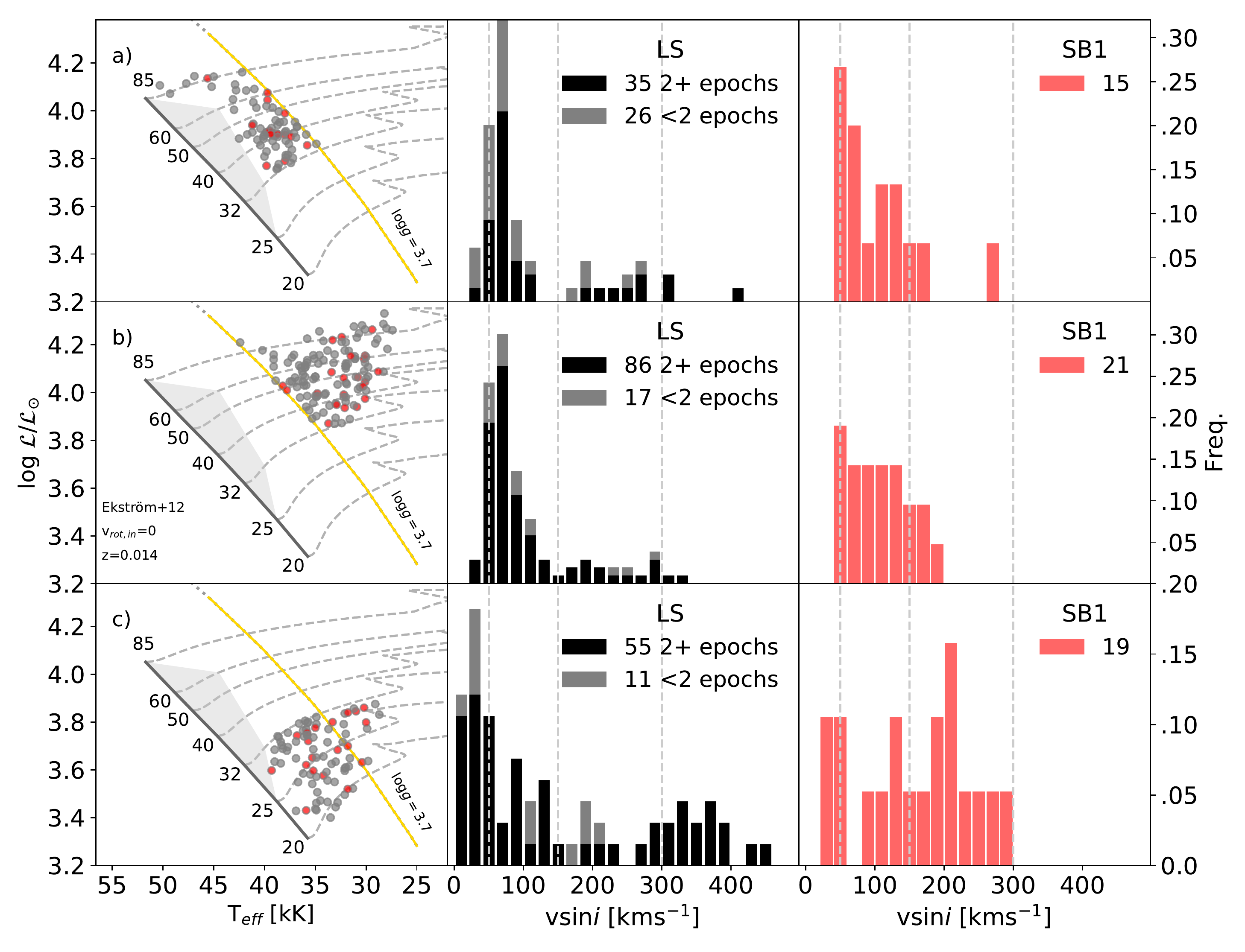}
\caption{Rotational properties of our working sample of Galactic O-type stars separated in three groups ({\em from top to bottom}), comprising stars covering different ranges in mass and evolutionary status. {\em [Left panels]} Location of each subgroup in the sHRD; see also Fig.~\ref{Gen_Sample_HR} for additional information about other features presented in each sHRD. {\em [Middle and right panels]} Corresponding \vsini\ distributions for the LS and SB1 samples, respectively, within each group. In each panel, we indicate the \vsini\ values of 50, 150, and 300~\kms\ with vertical dashed lines, for reference purposes. Similarly to Fig.~\ref{RotCompare_CSB1_VH}, the stars in the LS sample, for which we only count on one spectrum, are indicated in gray in the corresponding \vsini\ distributions.}
\label{Mezcla_histogr}
\end{figure*}

\subsection{\vsini\ dependence on mass and evolution}\label{vsini_evol}

In order to better investigate how the distribution of projected rotational velocities in the O-star domain depends on mass and evolution, we decided to go beyond the traditional use of the \Teff\,--\,\vsini\ diagram (Fig~\ref{Teff_vs_vsini}) and split the global sample of studied stars into three subsamples according to their location in the sHRD. By doing this, we could still keep a good level of statistical significance in each of these three subgroups, with a minimum of 75 objects in the less populated group. 
There have been similar studies for very large samples of lower-mass stars \citep[e.g.,][ for A-type stars]{Zorec2012}.

We first used the 32~\msol\ non-rotating evolutionary track computed by \cite{Ekstroem2012} to roughly separate the sample in two mass ranges. We chose this boundary because stars with higher masses are expected to develop winds strong enough to produce significant losses of angular momentum, even during the early phases of evolution along the main sequence. Then, we further divided the high-mass sample in two, using the \grav\,=\,3.7 dex line of constant gravity, which allowed us to roughly separate the expected younger dwarf stars from the more evolved giants and supergiants (see Fig.~\ref{Gen_Sample_HR}). 

The leftmost panels of Fig.~\ref{Mezcla_histogr} show the location of these three groups of stars in the sHRD, where we again highlight the stars identified as SB1s in red. These panels are complemented with the corresponding empirical \vsini\ distributions, separating once more the likely single and SB1 subsamples (middle and rightmost panels).

In each of these latter panels, we indicate  three \vsini\ values with vertical lines. These correspond to the rough boundary (50~\kms) below which our empirical \vsini\ measurements can be affected by some type of methodological limitations (see Sect.~\ref{lowvalues_intext}), the approximated limit we established in the global \vsini\ distribution between the low-velocity component and the tail of fast rotators (150~\kms, see Sect.~\ref{bimodalSect}), and the \vsini\ value above which no SB1 stars are detected (300~\kms, see Sect.~\ref{LSySB1}).

Hereafter, we denote the two high-mass subgroups as "a" and "b", respectively, while group "c" corresponds to the sample of O stars below the 32~\msol\ evolutionary track. In a single star evolution context, stars in group “a” can be regarded as progenitors of the stars in group “b”. Equivalently, the \vsini\ distribution of group “a” is expected to become the \vsini\ distribution of the group “b” by the effect of stellar evolution. Again in this context, group "c" could be considered as a separated group in terms of evolution. However, as illustrated by \cite{Ahumada2007}, \cite{Sana2012}, \cite{Schneider2014}, and \cite{Wang2020}, one must keep in mind that this naive interpretation of the location of stars in the sHRD in evolutionary terms becomes a bit more complex if we take into account that a high percentage of stars in our working sample might have suffered from some type of interaction with a companion, even if they are not presently detected as a spectroscopic binary. Therefore, in this section, we restrict ourselves -- as a first step -- to highlighting the main characteristics of the various \vsini\ distributions presented in Fig.~\ref{Mezcla_histogr} from a purely empirical point of view. We refer the reader to Sect.~\ref{sectionDiscuss} for a further discussion and interpretation of results taking into account our present theoretical knowledge about single and binary star evolution of high-mass stars.

Roughly speaking, it becomes clear from inspection of the various panels in Fig.~\ref{Mezcla_histogr} that the individual \vsini\ distributions of the LS and SB1 samples comprising each of the three subgroups retain similar properties to the corresponding ones resulting from the global sample (see Sect.~\ref{LSySB1}). In particular, we refer to the bimodal character of the \vsini\ distributions associated with the LS samples, and the more uniform distributions found in the cases of the SB1 samples.

To complement the information presented in Fig.~\ref{Mezcla_histogr} and help to better identify additional similarities and differences between the various \vsini\ distributions, we summarize some statistical quantities of interest, extracted from each of the distributions, in Table~\ref{meanvsini}. In all cases, we provide separated information for two \vsini\ ranges, using 150~\kms\ as boundary.

The first remarkable result is the striking similarity\footnote{A Kuiper's test indicates that there is a relatively high probability ($\sim$20\%) that these two samples are randomly produced from the same parent distribution.} of the \vsini\ distribution associated with the LS samples comprising groups "a" and "b". In both cases, there is a clear separation between the low- and fast-\vsini\ components at $\sim$150~\kms. In addition, the statistical properties of both components in the two groups (in terms of relative percentage of stars, associated mean \vsini\ value, and standard deviation, see Table~\ref{meanvsini}) are basically the same. Also, the corresponding \vsini\ distributions for the SB1 samples are remarkably similar (although with a much lower statistical significance), especially for the covered range in \vsini, extending from $\sim$50 to $\sim$200~\kms\ (except for one star in group "a", which is very close to the boundary with group "c").

\begin{table*}[!t]
        \caption{\vsini\ statistics for six star subsamples depicted in Fig.~\ref{Mezcla_histogr}.} 
        \label{meanvsini}
        \centering
        \begin{tabular}{cccccccccccccccccc}
                \hline
        \noalign{\smallskip}
               & \multicolumn{8}{c}{Sample with \vsini\,$<$\,150~\kms}   & & \multicolumn{8}{c}{Sample with \vsini\,$>$\,150~\kms}  \\
               \cline{2-9} \cline{11-18}
        \noalign{\smallskip}
               &       & \multicolumn{3}{c}{LS} &  & \multicolumn{3}{c}{SB1}  & &    & \multicolumn{3}{c}{LS} &  & \multicolumn{3}{c}{SB1}  \\
               \cline{3-5} \cline{7-9} \cline{12-14} \cline{16-18}
        \noalign{\smallskip}
               &  N$_{\rm all}$ &  N$_{\rm LS}$ &  \%  & $\langle$\vsini$\rangle$ & &  N$_{\rm SB1}$ & \% & $\langle$\vsini$\rangle$ & & N$_{\rm all}$  &  N$_{\rm LS}$ &  \%  & $\langle$\vsini$\rangle$ & &  N$_{\rm SB1}$ & \% & $\langle$\vsini$\rangle$ \\ 
               \cline{2-9} \cline{11-18}
        \noalign{\smallskip}
               a) & 59  & 47 & 80 & 65\,$\pm$\,19 & & 12 & 20   & 80\,$\pm$\,30 & &  17  & 14 & 82 & 252\,$\pm$\,65 & & 3 & 18  & 197\,$\pm$\,46   \\
               b) & 103 & 85 & 82 & 73\,$\pm$\,24 & & 18 & 18   & 92\,$\pm$\,30 & &  21 & 18 & 86 & 241\,$\pm$\,50 & & 3 & 14  & 171\,$\pm$\,9     \\
               c) & 50  & 41 & 82 & 53\,$\pm$\,38 & & 9 & 18   & 83\,$\pm$\,40 & &  35 & 25 & 71 & 300\,$\pm$\,82 & & 10 & 29   & 223\,$\pm$\,36   \\
                \hline
        \end{tabular}
\smallskip

{\tiny{\bf Notes:} extracted from the \vsini\ distribution associated with the six star subsamples depicted in Fig.~\ref{Mezcla_histogr} (LS and SB1 stars in a, b, and c panels), separated in two ranges of \vsini, comprising the low-velocity peak (\vsini\,<\,150~\kms) and the tail of fast rotators (\vsini\,>\,150~\kms). The columns labeled with $\langle$\vsini$\rangle$ quote the mean and standard deviation of the \vsini\ distribution within the indicated range. }
\end{table*}

This empirical result has far-reaching implications that will improve our understanding of the importance that angular momentum losses because of stellar winds, as well as several proposed mechanisms that could transport angular momentum from the stellar core to the envelope, have on the evolution of surface rotational velocities in stars with masses above 30~\msol\ (see Sect.~\ref{Lossingle}). In this regard, if we assume that there is an evolutionary connection between both groups, the abovementioned result would imply that no important braking is detected in any of the two main components of the \vsini\ distribution as stars evolve from region "a" to region "b".

Indeed, the mean \vsini\ value associated with the low-velocity component of the LS distributions is slightly higher in the case of group "b", something that is not expected from and evolutionary point of view. However, this result is most likely a consequence of the methodological limitation associated with our \vsini\ measurements. As commented in Sect.~\ref{lowvalues_intext}, this limitation may affect the relative number of stars populating the first two bins of the distributions. The lack of stars produces a shift in the associated mean \vsini\ value of the low-velocity component to somewhat larger values, especially in the case of stars in group "b".

Again (see also Sect. \ref{vsinisHR}), inspection of the \vsini\ distributions of stars in group "c" (and particularly that of the LS sample) allows us to confirm that the abovementioned methodological limitations do not significantly affect our \vsini\ estimates in the low-luminosity O-type stars . Contrarily to the case of the higher-mass samples, in which there is a very low percentage of stars with \vsini$<$40-50~\kms\,, the sum of the first two bins of the \vsini\ distribution of the LS sample comprising group "c" reaches almost 30\%. This feature, together with: (a) the existence of a clearly separated group of LS stars populating the \vsini\ range between $\sim$250 and 450~\kms\ in the lower-mass sample, which is not present at higher masses; and (b) the different \vsini\ limit above which no SB1 stars are detected in groups "c" ($\sim$300~\kms) and "a" ($\sim$200~\kms), seems to indicate that there is a dependence on mass in the efficiency of the various physical processes, which may affect the rotational properties of high-mass stars in the O-star domain along their main sequence evolution.

\section{Comparison with well-established theoretical predictions}\label{sectionDiscuss}

In this section, we provide further insights on the interpretation of the distribution of our working sample of O-type stars in the various diagrams presented in Sect.~\ref{sectionResults}. To this aim, as a guideline we use some of the theoretical predictions resulting from well-established studies that investigate the importance that single and binary star evolution can have on the rotational properties of massive stars \citep[e.g.,][]{Brott2011, Ekstroem2012, deMink2013,Wang2020}.
Being aware of the existence of other more recent model computations obtained with other single and binary stellar evolution codes, such as PARSEC \citep{Bressan2012}, FRANEC \citep{Degl'Innocenti2008}, BASTI \citep{Hidalgo2018}, or BPASS \citep{Eldridge2008}, we decided to mainly focus on the abovementioned computations because they are well established within the community and have been the object of previous comparisons.

As a disclaimer, we remark that we do not intend to tight-fit the empirical distributions using the available sets of single star evolutionary models and binary population synthesis simulations. Rather, we use the predicted trends to better understand the empirical properties of the sample under study and, at the same time, to evaluate possible observational constraints that our empirical results could provide to presently available (as well as future) theoretical developments.


\begin{figure*}[!ht]
\includegraphics[width=1.1\textwidth]{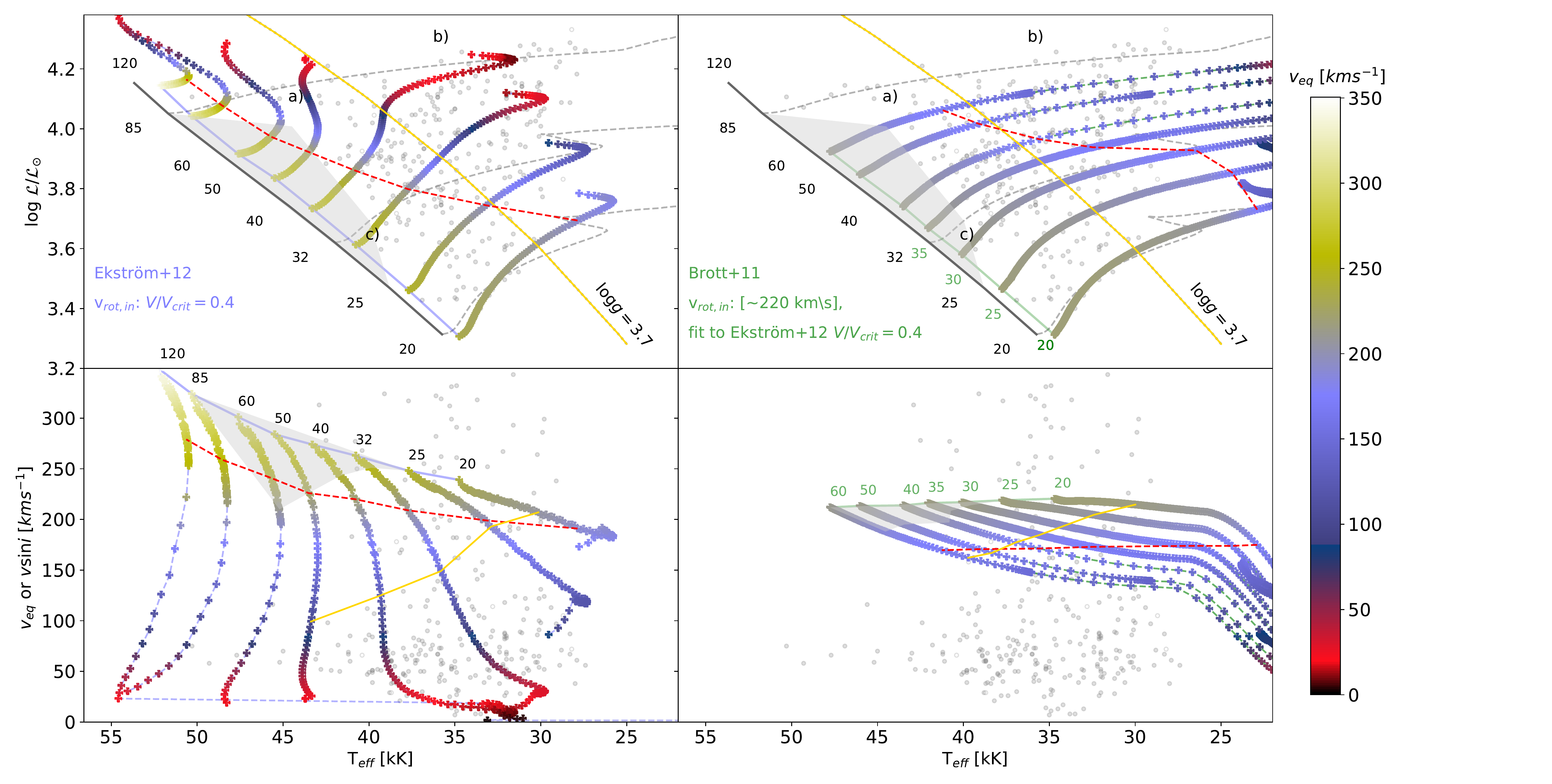}
\caption{Predicted behavior of the equatorial rotational velocity in the single star evolutionary models, with rotation computed by \citet[][{\em left panels}]{Ekstroem2012} and \citet[][{\em right panels}]{Brott2011}. See Sect.~\ref{Lossingle} for an explanation of the selected values of initial rotational velocity in each set of models. Cross symbols represent time steps in each evolutionary tracks, colored accordingly to the corresponding \veq. The dashed red line marks the point in each track with a 20\% reduction of \veq\ from \vini, and the yellow line indicates the \grav\,=\,3.7~dex line of constant gravity. Observations are represented by gray points. The shaded gray area highlights the area in the sHRD close to the ZAMS, which is void of observed stars. {\em [Top panels]} Considered evolutionary tracks in the sHRD, where we also depict the ZAMS and the 20, 32, and 85~\msol\ evolutionary tracks corresponding to Ekstr\"om+12 models without rotation to better identify the three regions in the sHRD (marked as "a", "b", and "c", respectively) discussed in Sects.~\ref{vsini_evol} and \ref{Lossingle} (see also Figs.~\ref{Mezcla_histogr} and \ref{Regions_veq}). {\em [Bottom panels]} Considered evolutionary tracks in the \veq\,-\,\Teff\ diagram (or \vsini\,-\,\Teff\ for the case of the observations).}
\label{EvolModsHR}
\end{figure*}


\begin{figure*}[!ht]
\includegraphics[width=\textwidth]{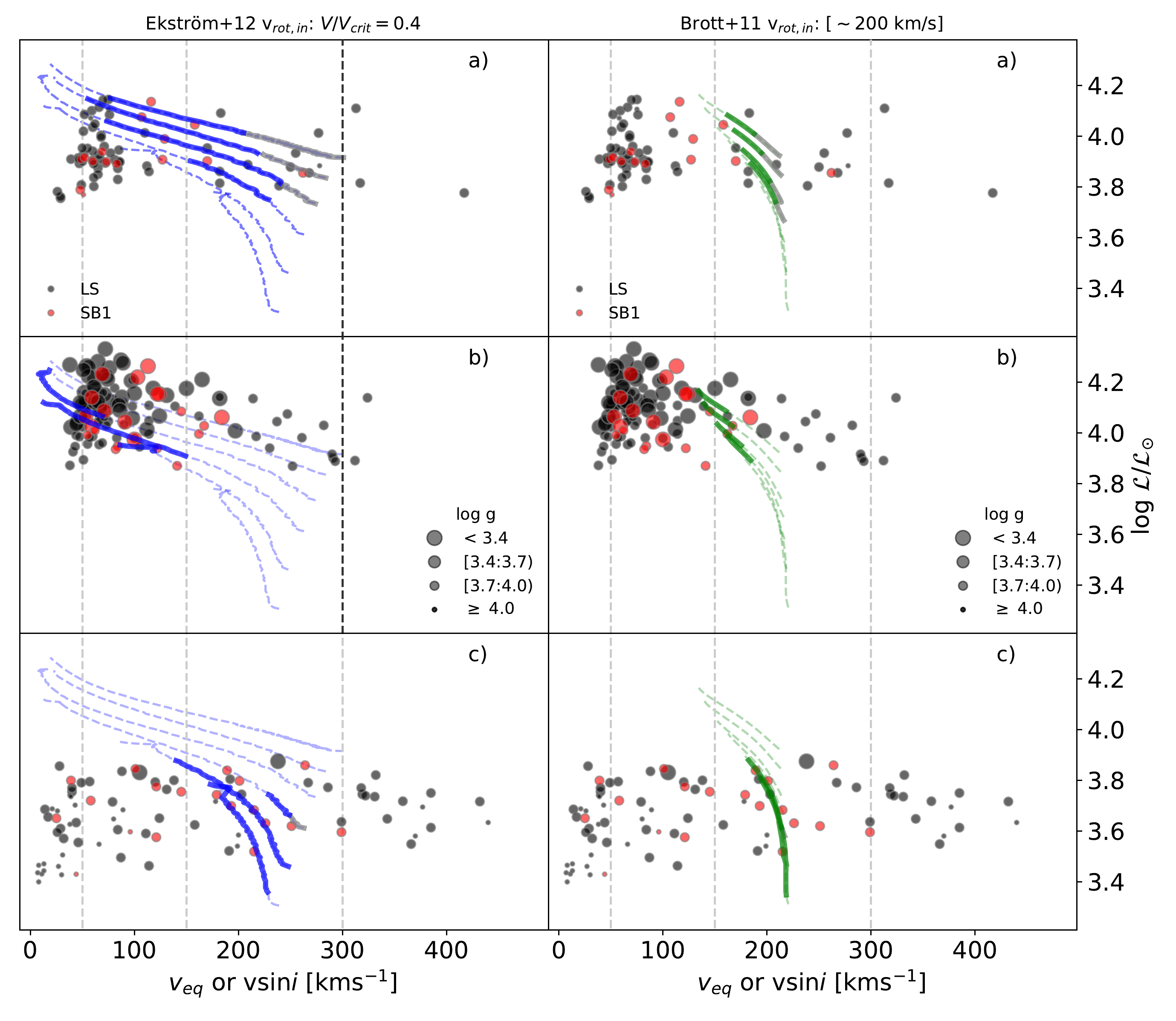}
\caption{Predicted behavior of the evolutionary tracks computed by \citet[][left panels]{Ekstroem2012} and \citet[][right panels]{Brott2011} in the \veq\,--\,\Lsp\ diagram. From top to bottom, the sections of the evolutionary tracks, which are located within the three regions ("a", "b", and "c", respectively) indicated in Fig.~\ref{Mezcla_histogr} and described in Sect.~\ref{vsini_evol}, are highlighted with blue and green thicker lines. In addition, the gray thick part of the evolutionary tracks indicates that the corresponding section of the track lies inside the region of the sHRD close to the ZAMS where no O stars are detected.
The stars in the observed sample that populate the three considered regions of the sHRD are also included in each panel, separately, for comparison (this time using \vsini\ values instead of \veq). As in other figures, we indicate three \vsini\ values of interest (50, 150, and 300~\kms) with vertical lines. Size represents \grav\ value. 
}
\label{Regions_veq}
\end{figure*}

\subsection{Single star evolutionary models}\label{Lossingle}

Figure~\ref{EvolModsHR} shows the behavior of the two sets of single star evolutionary models including the effect of rotation that have been commonly used by the massive star community in the last decade. Namely, those of \cite{Brott2011} and \cite{Ekstroem2012}, hereafter quoted as the Brott+11 and Ekst\"om+12 models. In particular, for the case of the latter models, we chose those computed with an initial rotational velocity (\vini) being 40\% of the critical velocity (\vcrit) at the ZAMS. For consistency, among the models available in the Brott+11 database\footnote{The Brott+11 database includes models computed in a wide range of \vini, from zero to 600 \kms.},  we selected those assuming a \vini\,$\sim$\,220~\kms.

The top panels in Fig.~\ref{EvolModsHR} present the evolutionary tracks of both set of models in the sHRD. For reference purposes -- and to better identify the three regions in the sHRD discussed in Sect.~\ref{vsini_evol} (see also Fig.~\ref{Mezcla_histogr}) -- we also depict the \grav\,=\,3.7~dex line of constant gravity, and the ZAMS and the 20, 32, and 85~\msol\ evolutionary tracks corresponding to the Ekstr\"om+12 models without rotation.
The bottom panels in the same figure show the predicted evolution of equatorial velocity (\veq, without inclination effect), using \Teff\ as a proxy of time (see also the corresponding empirical diagram in Fig.~\ref{Teff_vs_vsini}).

In all panels, tracks are colored depending on the value of \veq, and the dashed red line marks the point on each track where  \veq\ has dropped by 20\% of its \vini. We also highlight, as a gray shadowed area, the region in the sHRD close to the ZAMS, which is void of observed O stars \citep{Holgado2020}. Knowing the specific location of this region in the \veq\,--\,\Teff\ diagram is of critical interest when we compare model predictions and the distribution of observational points. If this particular region of the \veq\,--\,\Teff\ diagram is void of stars, it could mean that either stars with these properties do not exist, or they have escaped from our optical spectroscopic surveys.

As expected, Fig.~\ref{EvolModsHR} shows that the models for single massive stars predict a braking effect that is stronger for the most luminous stars because of stronger winds and enhanced mass loss \citep{Brott2011, Ekstroem2012}. It also shows that the Ekstr\"om+12 models slow down faster than the Brott+11 models, the effect being more clearly seen in the lower panels. This is mainly a consequence of the specific treatment of an internal magnetic field in each of the two evolutionary model computations, which in turn causes the consequent reinforcing and/or limitation of angular momentum transport within the star \citep[see][and references therein]{Martins2013, Keszthelyi2017, Keszthelyi2020}.

The model predictions presented in Fig.~\ref{EvolModsHR} are also complemented with the corresponding distribution of empirical points in each of the considered diagrams. From first inspection of the lower panels of this figure, it becomes already clear that -- although the \veq\,--\,\Teff\ (or \vsini\,--\,\Teff) diagram has been used elsewhere to contrast model predictions about the spin evolution of massive stars with observational data -- the comparison of empirical and modeled values using exclusively this diagram is complex. In particular, we have to take into account that stars with a given \Teff\ and different luminosities (and thus masses) can appear mixed in this diagram, and are not necessarily placed near their correct corresponding evolutionary track.

Neither of the two sets of evolutionary models with rotation is able to reproduce the lack of stars with \vsini\ values between $\sim$75~\kms\ and the considered \vini\ at the hot side of the diagrams, an effect that is more evident in the Ekstr\"om+12 models. Moreover, Brott+11 models with \vini$\sim$200~\kms\ cannot explain the existence of low-velocity stars across the whole O-star domain, as they do not slow down sufficiently before reaching temperatures below 30\,000 K. 

These results seems to indicate that single star evolutionary models with an initial equatorial velocity of $\lessapprox$150~\kms\ might be more convenient to explain the spin-rate properties of a high percentage (75\%, as stated in Sect.~\ref{bimodalSect}) of the stars in our sample. However, this consideration will still leave us with an unsatisfactory explanation for the stars
with rotational velocities above the selected \vini.

In Fig.~\ref{Regions_veq} we show the distribution of empirical points in the \veq\,--\,\Lsp\ diagram, along with the predicted behavior of the evolutionary tracks for the same set of evolutionary models considered for Fig.~\ref{EvolModsHR}. For a better comparison, we present a separated diagram for each observational subsample on the three areas of the sHRD defined in Fig.~\ref{Mezcla_histogr} (from top to bottom), as well as for the Ekstr\"om+12 (left) and Brott+11 (right) models.
In each panel, we highlight the section of the evolutionary tracks that corresponds to the represented area with a thicker colored line. This is the part of the evolutionary track that has to be compared with the subsample of observational points depicted in each panel. The other parts of the tracks, represented as thin dashed lines, are only included for reference; they do not cross the corresponding area in the sHRD (see Fig.~\ref{EvolModsHR}).

As in previous figures, we also mark the sections of the evolutionary tracks that correspond to the area near the ZAMS void of stars with a thick gray line.
Lastly, we note that in this figure, the temporal evolution goes from right to left, and we recall that the quantity represented in the x-axis is different for the model (\veq) and the observations (\vsini); thus, the projection effect must be taken into account for a correct interpretation of the information presented in this figure.

At first glance, the empirical distributions do not coincide with the expected theoretical ranges in most cases.
In panels (a) and (b) -- namely stars with M$>$32~\msol\ at gravities higher and lower than \grav\,=\,3.7~dex, respectively -- most of the observational points appear at rotational velocities much lower than those predicted by the Brott+11 models (right column). Although there is a projection effect involved, the high number of points clustering around \vsini$\sim$65~\kms\ (see also Fig.~\ref{Mezcla_histogr} and Table~\ref{meanvsini}), allows us to discard this effect as a possible explanation of the discrepancy. This result reinforces our previous statement that -- if interpreted in terms of Brott+11 model predictions -- an initial velocity of $\sim$150~\kms\ (or even lower) seems to be more appropriate to explain the rotational properties of more than 70\% of the O stars in the high-mass region.

The situation is somewhat different -- but still not very encouraging -- for the case of Ekstr\"om+12 models (left panels). As previously commented (see also bottom left panel in Fig.~\ref{EvolModsHR}), the specific physical ingredients considered by \cite{Ekstroem2012} in their evolutionary model computations lead to an important braking of the stellar surface soon after the star leaves the ZAMS, especially in the case of stars with masses above $\sim$30~\msol. In addition, as illustrated by the top left panel in Fig.~\ref{EvolModsHR}, the Ekstr\"om+12 evolutionary tracks depart from their non-rotating counterparts, this effect becoming more important for increasing masses. Both effects together make it possible to explain the existence of stars in region "a" with \Lsp\ in the range $\sim$4.0\,--\,4.2~dex, with both low and intermediate values of \vsini\ (even if models with \vini/\vcrit\,=\,0.4 are considered). Also, the 32 and 40~\msol\ evolutionary tracks reach region "b" with equatorial velocities that are compatible with the low-\vsini\ component of the empirical distribution of points. However, there is still a large amount of stars in regions "a" and "b" with a location in the \veq\,--\,\Lsp\ diagram that is incompatible with the considered Ekstr\"om+12 models.

The top and middle panels of Fig.~\ref{Regions_veq} also allow us to contrast which of the two model predictions regarding the spin evolution of stars in the mass range between $\sim$30 and $\sim$80~\msol\ is in better agreement with the observations.  As already commented in Sect.~\ref{vsini_evol} (and illustrated by Fig.~\ref{Mezcla_histogr}), the relative percentages of stars in regions "a" and "b" with intermediate values of \vsini\ (50\,-\,100~\kms) are quite similar. This is in contrast with the strong braking of the stellar surface predicted by the Ekstr\"om+12 models, but in relative fair agreement with the more moderated spin down characterizing the Brott+11 models.

In this context, we must remember that there is empirical evidence indicating that our line-broadening analysis strategy might be failing for some stars for which we obtained \vsini\ estimates between $\sim$40\,--\,50~\kms\ (see Sects.~\ref{lowvalues_intext} and \ref{vsini_evol_Teff}); however, as illustrated by Fig.~\ref{logLsp_vs_Vsini}, this methodological limitation should affect the stars in the low- \vsini\ boundary of the distribution in both regions "a" and "b"in a similar way, and thus we expect the abovementioned statement to still be valid once we are able to overcome this methodological issue. 

Furthermore, the strong braking effect present in Ekstr\"om+12 models makes it challenging to find a straightforward explanation -- even if we assume a higher value of \vini/\vcrit\ -- for the existence of evolved stars (i.e., those in region "b") with \Lsp$\gtrsim$4.0 and \vsini\ values in the range 50\,--\,300~\kms. Indeed, this conclusion is not affected by the limitations of our line-broadening analysis strategy since we are referring to \vsini\ estimations well above the problematic range.

This latter result could be considered as a strong empirical argument to support the various decisions taken by \cite{Brott2011} regarding the treatment of the internal angular momentum transport in the evolutionary model computations. In particular, the activation of the effect of internal magnetic fields (via the Taylor-Spruit dynamo) on this transport helps to compensate the braking of the stellar envelope produced by the presence of strong stellar winds in the more massive stars.

However, we must emphasize that this is not necessarily the last word on this matter, since the mass loss recipes assumed in the model computations also have important consequences for the predicted evolution of surface angular momentum in massive stars \citep{Keszthelyi2017}. For example, these authors showed that, decreasing the usually considered mass loss rates by 30\% \citep[from][]{Vink2000}, they could make a model without an internal magnetic field result in a similar evolution of \veq\  as a model with a magnetic field with the original values of mass-loss rate. Interestingly,  
\cite{Bjorklund2022} has recently provided a new prescription for steady-state, radiation-driven mass loss from hot, massive stars, also indicating that for the case of O stars, the predicted mass loss rates are lower by about a factor  of three than the rates typically used in previous stellar evolution calculations. It would therefore be interesting to perform a similar comparison to the one presented in Fig.~\ref{Regions_veq}, but assuming a 30\% of reduction in the mass loss rates incorporated in the Ekstr\"om+12 models (and a smaller value of \vini/\vcrit, see Sect.~\ref{ViniSect}). 

This problem of reconciling observations and models was already pointed out by different previous studies. 
\cite{Markova2018}, using 53 Galactic O-type stars and the same sets of models, found that none were able to reproduce the helium and nitrogen surface enrichment appropriately when considering only the expected behavior of the \vsini\ evolution included in single star rotating evolutionary models.  
Also, \cite{Martins2017} found difficulties in explaining the chemical properties of 15 late O giants at low metallicity using the single star evolution models.

Regarding stars in region "c", the main problem arises from the fact that neither of the two sets of models is able to explain the broad range in \vsini\ that characterizes the empirical distribution. In this case, angular momentum losses due to stellar winds are much less relevant, and thus the only way to be able to reconcile single star evolutionary model predictions with the empirical distribution of points would be to consider that stars in this mass range are born covering a very broad range of initial spin rates (from almost zero and up to $\sim$60\% of the critical velocity at the ZAMS, see Fig.~\ref{vsiniage0}). Indeed, to some extent, a similar problem is also present in the upper mass region, especially if we assume that angular momentum losses because of stellar winds do not result in an important braking effect of the stellar envelope (and thus the observed values of \vsini). 

In view of the discussion above, one would be tempted to explain the various features characterizing the observational distributions presented in Figs.~\ref{Teff_vs_vsini}, \ref{logLsp_vs_Vsini}, and \ref{Regions_veq} by means of a combination of Brott+11 models with different \vini. However, this would result in a very ad hoc distribution of initial rotational velocities. For example, an anticorrelation between the maximum initial rotational velocity and the initial mass would be needed, in order to reproduce the lack of massive stars with high rotational velocities close to the ZAMS and the large number of lower-mass stars with high rotational velocities. In addition, as already commented, in the 20\,--\,40\msol\ range, we would need a much broader range of initial spin rates than in the case of stars with higher masses (see also further notes in Sect.~\ref{ViniSect}). Last but not least, there are some features in the distributions that could be only explained by assuming that some stars are spinning up along their evolution, something that is not possible in the case of single star evolution.

\subsection{Exploring the effect of binary interactions}\label{BinInteract} 

The present formation scenario of massive stars states that $\sim$90\% of them are born as binary\footnote{Or even, a non-negligible percentage of them, as part of a hierarchical multiple system \citep[][]{MaizApellaniz2019,Trigueros2021}.} systems \citep{Sana2012}, As a result, it is natural to think that the features in the various diagrams presented in previous sections that cannot be easily explained by means of single star evolutionary models may be produced by binarity effects. To evaluate this, we follow the ideas presented by \cite{Ramirez-Agudelo2015} in their study of the rotational properties of O-star binary systems in the 30~Doradus regions of the LMC, which, ultimately, are based on theoretical results obtained by \cite{deMink2013}.

The evolution of the individual spin rates in a binary system can be
affected by several additional effects not present in the case of isolated stars, some of them giving rise to a more or less important
acceleration of one of the two components (or even both under certain circumstances). 
For example, in close binaries (P\,$\lesssim$\,10~days), tidal interaction may synchronize rotation and orbit\footnote{In wider orbit binaries, tidal effects can be considered as negligible. Indeed, it has been claimed that it is the stellar formation process, rather than tidal evolution, that can be at the origin of any plausible differences (up to a factor of five) on the rotation of their components \citep[see][and references therein]{Putkuri2021}.}  
(either increasing or decreasing the equatorial velocities). But more importantly in the context of our study, mass (and angular momentum) transfer occurring after Roche-lobe overflow can spin up the initially less massive star to a very high rotational velocity, which can even reach critical velocity in some cases \citep[see][and references therein]{Petrovic2005, deMink2013}. In this situation, the secondary star in the binary system does not only experience an increase in rotational speed, but it also modifies its luminosity and apparent age. Eventually, this mass-transfer event could modify the appearance of the binary system, leading to a fast-rotating, rejuvenated (apparently) single object in many cases \citep{Ahumada2007,Schneider2014,Schneider2015,Schneider2016,Wang2020}. 
All of this simulated behavior should be considered with caution, since it relies on some physics that are not fully constrained. There are still different criteria when considering the amount of mass and angular momentum that is lost in collisions and mass transfer events \citep{Wang2022}, as well as in the generation of a magnetic field and its braking properties in mergers events \citep{Schneider2019}.

The binary population synthesis simulations performed by \cite{deMink2013} showed that binary interaction offers a natural explanation for the bimodal characteristic of the global rotational distribution commonly found in the O-star domain (Sect.~\ref{bimodalSect}). By assuming a flat initial rotational distribution between 0 and 200~\kms\ -- and a continuous star formation -- they found that, after several Myrs, the effects of evolution, tides, mass transfer, and mergers modify this initial distribution to another one, including a dominant low-velocity component and a tail of fast rotators. In this context, the simulations by \citeauthor{deMink2013} predict that 19\% and 11\% of the stars will develop rotational speeds above 200 and 300~\kms, respectively. These numbers can be compared with the 17\% and 7\% resulting from our empirical distribution of projected rotational velocities.
In this context, we find it interesting to indicate that the total percentage of detected spectroscopic binaries in our sample of 415 stars is $\sim$41\% (55 SB1s and 113 SB2s). Indeed, it could reach $\sim$50\% if we roughly consider half of the LS stars with a single spectrum as hidden binaries. These numbers must be taken into account when comparing with the simulation from \citeauthor{deMink2013}, who assumed an initial binary fraction of 70\%. 
That would imply that about 20\% of the present single stars should have been born as binaries.
Also, more in-depth future studies connecting the complete evolutionary path followed by massive stars should take into account the increasingly available empirical information about detected spectroscopic binaries in more evolved stages, such as B supergiants \citep[][and in prep.]{SimonDiaz2020b}, red supergiants \citep[e.g.,][]{Neugent2021,Dorda2021}, or WR stars \citep[e.g.,][]{Dsilva2022}, which routinely obtain much lower binary fractions.

Aside from projection effects, we can expect some differences between our empirical results and de Mink's simulations due to processes favoring a faster spin down than modeled, such as the formation of magnetic fields during mass accretion \citep[as mentioned by][]{deMink2013}. Evidence supporting such processes (not necessarily a magnetic field) is found in the observation of super-synchronization but clearly subcritical rotation in OB+WR systems \citep{Shara2020}. Moreover, we note that the \cite{deMink2013} simulation includes SB2 stars that have not been considered in our observational sample, and that the difference in metallicity -- LMC for de Mink's study -- could lead to a higher number of high-spin objects in their simulations. Considering all these uncertainties, the agreement seems to be surprisingly good. The scenario proposed by \cite{deMink2013} provides a more natural way to explain the distribution of stars in the \vsini\,--\,\Teff\ and \vsini\,--\,\Lsp\ diagrams (Figs.~\ref{EvolModsHR} and \ref{Regions_veq}, respectively) than the assumption of a very ad hoc initial spin-rate distribution (see Sect.~\ref{Lossingle}).
Despite considering that the most straightforward solution for bimodality seems to be the effect of binary interaction and its spin-up effect on stars, it is not possible to totally rule out the existence of stars that are born with an extremely high rotation naturally, although as assumed in Sect.~\ref{Lossingle}, it may seem a more artificial solution.

The second empirical feature that we evaluate is the clear dominance of apparently single stars in the tail of fast rotators (especially above 250\,--\,300~\kms\ in the lower-mass domain, and 150\,--\,200~\kms, for higher masses, see Fig~\ref{Mezcla_histogr}). This result agrees with the hypothesis that the tail of fast rotators is mostly populated by post-interaction products of binaries \citep{deMink2013} that, in addition -- as indicated by \cite{deMink2011, deMink2014} --  will often appear to be single because the companion tends to be a low-mass, low-luminosity star in a wide orbit. Alternatively, they became single stars after a merger or disruption of the binary system during the supernova explosion of the primary. 

On a related matter, we consider, as a third characteristic to be interpreted in the context of binary evolution, the apparent upper limit in \vsini\ of $\sim$300~\kms\ found in the sample of clearly detected SB1 stars (see Fig.~\ref{RotCompare_CSB1_VH}).\ This phenomenon is also present, although not as abrupt, in the study by \cite{Ramirez-Agudelo2015}, in spite of the different metallicity. 

In principle, our methodology for distinguishing between LS and SB1 stars (see Appendix~\ref{quan_var}) could be dismissed as a plausible factor, making this results to be spurious. Being aware that the detection of small changes in \vrad\ due to orbital motions becomes more complex when the rotational broadening becomes larger, we see no obvious reason indicating why the value of \vsini\,=\,300~\kms\ could represent a limitation. Indeed, it is also interesting to note that, although we identify some SB1 stars among the lower-mass sample (see Fig.~\ref{Mezcla_histogr}) with \vsini\ in the range $\sim$200\,--\,300~\kms, there is only one clearly detected SB1 star with \vsini\ in this range in regions "a" and "b".

We could also claim that this result is just a consequence of the limited number of epochs available for the sample of fast rotators. However, as indicated in Table~\ref{tableValues_over300}, we count on a minimum of five epochs for most of them (except for two stars)\footnote{See also a follow-up study of the full sample of O stars with \vsini\,$>$200~\kms\ (Britavskyi et al., in prep.) in which we investigate in more detail the detected line-profile variability also using a larger number of epochs for some of the stars (gathered with the STELLA telescope).}. 

Following \cite{deMink2014}, our SB1 star sample can be assumed to be dominated by pre-interaction objects - in other words, previous to mass exchange -- so the only phenomenon that may be altering the rotational velocity beyond that expected by single star evolutionary models is tidal interactions.
If we look at the results provided by \cite{deMink2013}, these tidal interactions have several visible effects on their simulated distribution, such as a tail of accelerated stars that reaches just 300~\kms. Hence, the effect of tidal interactions could explain why the maximum speed reached by our sample of clearly detected SB1 stars, if mostly pre-interaction, is precisely that value.

The effect of tidal interactions and orbital-rotation synchronization may also be the origin of the difference between the distributions of single and SB1 stars (Fig.~\ref{RotCompare_CSB1_VH}), the latter being characterized by a wider low-\vsini\ peak and a somewhat higher mean value of \vsini\ when compared to the former (see Sect.~\ref{LSySB1}). 
Furthermore, \cite{deMink2013} also indicate that tidal effects prevent stars in a binary system from slowing down excessively, something which could also explain a lower frequency of stars with \vsini$<$40-50\kms\ found among the SB1 sample (compared to the LS sample). 

All these three characteristics are also found in the work by \cite{Ramirez-Agudelo2013, Ramirez-Agudelo2015}, although with small differences (e.g., they have higher \vsini\ velocities, also for binary components) and without considering stellar masses and effective temperatures. These authors delved into the empirical existence of the bimodal distribution of \vsini\ and the differences between apparent distributions of single and binary components (mainly primaries). They also used the works by \cite{deMink2013, deMink2014} for the interpretation of their results, finding good agreement between observations and simulations. Our study reproduces their findings. Interestingly, however, in comparison with Ramirez-Agudelo's study, ours represents not only a different metallicity but also a different environmental origin, as the LMC sample comes from a single cluster, 30 Dor. This reinforces that the origin of the bimodality and the lower limit in \vsini\ among SB1 as compared to apparent single stars is, at least in the first order, independent of metallicity and environment.

The fourth and last characteristic we want to investigate is the specific location in the sHRD of those stars in the sample with \vsini\,$>$\,300~\kms\ (Sect.~\ref{vsini_evol} and Fig.~\ref{Mezcla_histogr}).
As already discussed, the origin of such a high velocity is probably the binary interaction . Mass accretion or merger during the evolution of the binary systems may produce a rejuvenation and spin up of the initial secondary star or the final product \citep[if no additional mechanisms such as the formation of magnetic fields play a role,][]{deMink2013, deMink2014}. They may even appear as blue stragglers in a cluster when compared to the rest of the population \citep{Ahumada2007, Schneider2014,Schneider2015,Schneider2016,Wang2020}. The works of \cite{Ahumada2007} and \cite{Schneider2016} find that the number of blue stragglers is larger, and their rejuvenation more intense, for smaller masses and more evolved clusters. This, together with a faster evolution at higher masses, could explain the concentration of very fast rotators that we see below 32 \msol.

\section{A final note on the initial spin-rate distribution}\label{ViniSect}



\begin{figure*}
\includegraphics[width=\textwidth]{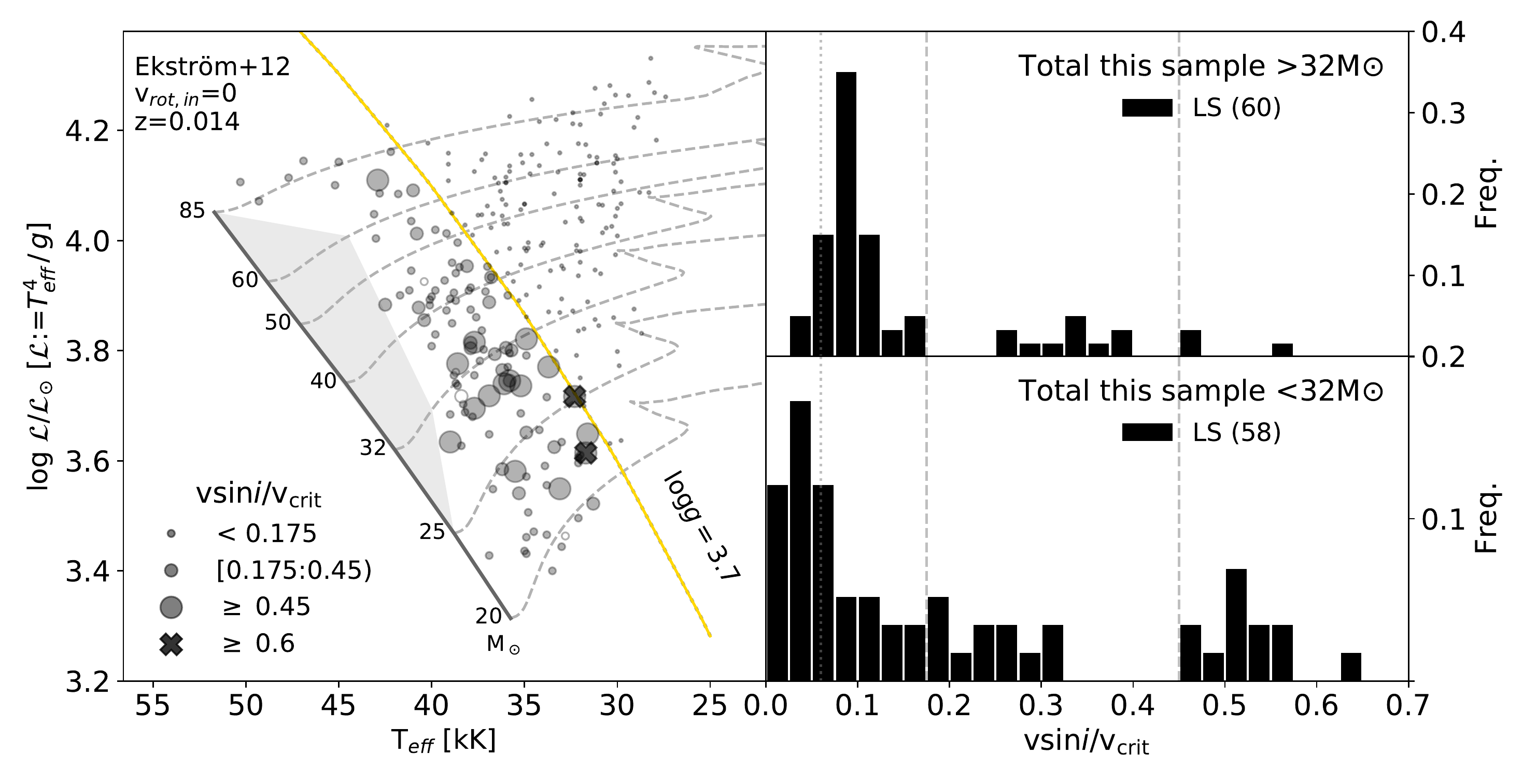}
\caption{Properties of the supposed youngest stars in the sample. (\textit{Left}) sHRD highlighting a subsample of 118 LS O-type stars with \grav\,$>$\,3.7~dex. The symbol size depends on the value of the ratio \vsini/\vcrit, as indicated. \textit{(Right)} \vsini/\vcrit\ distributions (as normalized histograms) for the indicated subsample, separating those stars located above (upper panel) and below (lower panel) the 32\msol\ track. The vertical dashed lines mark several values of interest for the \vsini/\vcrit\ ratio (see text for explanation). The vertical dotted line at \vsini/\vcrit=0.06 marks the approximate limit below which the \vsini\ determination can be affected by methodological limitations in those stars located above the 32~\msol\ track).
}
\label{vsiniage0}
\end{figure*}

The ratio of initial equatorial velocity (\vini) against the critical velocity
(\vcrit) is an influential input parameter for any evolutionary model of massive stars \citep[see review by][and references therein]{Maeder2000}. As a consequence, having access to reliable empirical knowledge about the initial spin-rate distribution of main sequence O- and B-type stars is of utmost importance for understanding the importance of rotation on the evolution and final fate of these important stellar objects. In addition, it provides a key constraint for our theories of massive star formation, and is an essential ingredient in population synthesis models.

There are several empirical studies in the literature that focus on the study of  the main characteristics of the \vini/\vcrit\ distribution associated with stars covering different mass ranges in the high-mass domain and located in different environments \citep[e.g.,][]{Wolff2006,Martayan2006, Huang2010}.
Of particular interest for comparison with our study are the works performed by \cite{Wolff2006} and \cite{Markova2014}. In their investigation of the spin-rate properties of a sample of newly formed stars spanning a range in mass between 0.2 and 50~\msol, \cite{Wolff2006} included 44 O-type stars. These authors, who define \vobs\ as the projected rotational velocity corrected statistically from the projection effect, find that the average ratio \vobs/\vcrit\ for stars between\footnote{Although they include two stars above 50 \msol\ in their sample, this number is too low for statistics} 25 and 50 \msol\ is 0.2, and 0.13 for stars between 8 and 25 \msol. We should note, however, that in their study, \citeauthor{Wolff2006} (a) neglected the effect of radiation pressure on the estimation of \vcrit\ (that may be important for the first group), (b) did not consider possible binarity effects, (c) did not make any attempt to disentangle the effect of macroturbulent broadening, and (d) as commented above, did a statistical correction to account for the effect of the inclination angle. 

More recently, \cite{Markova2014} extended the investigated mass range (now also disentangling the effect of macroturbulence on the line-broadening analysis) by studying 31 apparently single O stars, with 18 of them having initial masses between 50 and 80~\msol. They find that above 50 \msol\ the ratio \vsini/\vcrit\ does not exceed 0.26 (and is mostly below 0.17), whereas they find a limit of 0.50 between 22 and 50~\msol, with most stars rotating at less than 30$\%$ of their critical velocity. Indeed, only two stars in their sample exceed 200 \kms. 

The sample analyzed in this paper allows us to review and considerably extend the works performed by \cite{Wolff2006} and \cite{Markova2014} thanks to, among other things, the more complete mass range coverage, and the identification of SB1 stars. 

To construct an estimated \vsini/\vcrit\ distribution\footnote{Following \cite{Brott2011}, we use the definition \vcrit\,=\, $\sqrt{gR_{*}(1-\Gamma)}$, being $\Gamma$ the Eddington parameter for electron scattering. However, see also notes in \cite{Ekstroem2012} and \cite{Keszthelyi2017}.} from our data, we need to select the \vsini\ of the stars that have evolved as little as possible since birth. To this aim, we selected stars in our sample having \grav\,$>$\,3.7~dex, using gravity again as a proxy for evolution.
In Sect.~\ref{vsini_evol} we showed that the \vsini\ distribution tends to maintain shape during evolution (see also top and middle panels in Fig.~\ref{Mezcla_histogr}). This further reinforces our consideration of selecting stars up to the \grav\,=\,3.7~dex line of constant gravity to approach an initial rotation distribution while trying to keep statistics as large as possible in the resulting distribution. 
Lastly, we discard SB1 stars, as these have most likely suffered variations in their \vini\ since the beginning of their evolution because of the influence of their companions (see Sect.~\ref{BinInteract}). 

Although the \vsini\ values of the subsample described above should approximate in some way the initial velocity distribution of Galactic O stars, we must keep in mind that -- taking into account the ideas presented in Sect.~\ref{BinInteract} -- its similarity with the real distribution will largely depend on the net effect of the binaries on the sample, and the absence of a strong and fast braking effect from the ZAMS to the position of these stars.

The left panel of Fig.~\ref{vsiniage0} presents the location of the abovementioned subsample in the sHRD, whereas the right panel shows the associated \vsini/\vcrit\ histograms, further separating the investigated subsample in two (considering the stars located above and below the evolutionary track of 32 \msol). 

Both histograms show a peak near \vsini/\vcrit= 0.1, with the lower-mass stars extending to lower values (probably as consequence of the limitations in obtaining low \vsini\ values for stars of high luminosity, see Sects.~\ref{lowvalues_intext}, \ref{vsini_evol_Teff} and Appendix.~\ref{lowvalues}). But the most obvious feature in the two histograms is the existence of low- and high-velocity components clearly separated by a gap (between $\sim$0.175 and 0.25 for stars with M\,$>$\,32~\msol, and between 0.325 and 0.45 for stars below this mass. 

Following the discussion in Sect.~\ref{BinInteract}, we associate these gaps to the effect of binary evolution. On the one hand, the region with \vsini/\vcrit\,$<$\,0.175 would be dominated in both cases by single stars or wide binaries that have suffered little evolution, and thus are representative of the initial spin-rate distribution. On the other hand, following \cite{deMink2013, deMink2014}, the fast rotating components are most likely populated by post-interacting binaries. Therefore, these high-velocity components, which indeed comprise a small percentage of the investigated sample, should not be associated with the initial spin-rate conditions of stars at birth.

The difference in the upper boundary of the low-velocity components (and thus{ }the location of the gap) in the high- and low-mass distributions may have different interpretations. One possibility would be to attribute it to the influence of the stellar winds. Surface angular momentum losses because of stellar winds are expected to play a more important role in the higher-mass domain, thus producing some additional braking when compared to lower-luminosity stars. However, we recall that, in Sect.~\ref{Lossingle}, we showed that stellar winds do not seem to be very effective at producing an important braking of the stellar envelope as the more luminous O-type stars evolve. It could also be due to a small mass dependence of the covered range of the \vini/\vcrit\ ratio, with decreasing values for increasing initial stellar mass. This would be the opposite effect to the one presented by \cite{Wolff2006}, but it agrees with the results by \cite{Huang2010} for 220 high-gravity B stars. They find  a decreasing range of \veq/\vcrit\ from 0.8 for $M$\,=\,2~\msol\ stars to 0.2 on the frontier with O stars. Lastly, the low-mass sample could still have a small contamination of SB1 stars between \vsini/\vcrit\,$\sim$\,0.2\,--\,0.3, as a significant number of stars in this range have less than two epochs (see Fig.~\ref{Mezcla_histogr}).

Whatever the reason for the differences between the two \vsini/\vini\ distributions, and even considering projection effects, we conclude that O stars 
in their unaltered early stages have velocities between \vsini/\vcrit\,$\sim$\,0.1-0.175.
If we rule out a fast and extreme braking effect just next to the ZAMS, not present in any of the two sets of evolutionary models considered here, O-type stars are typically born with rotational velocities near \vini/\vcrit\,$\sim$\,0.1 and not significantly exceeding \vsini/\vcrit\,=\,0.2 \citep[in agreement with the results presented by][]{Markova2014}. Therefore, evolutionary models for single stars should not use higher values (except perhaps for some peculiar individual cases). Stars in binary systems will be spun up in the course of their evolution through binary interaction, populating the upper \vini/\vcrit\ values up to ratios $\leq$0.7.

\section{Summary and conclusions}\label{Sect_summary}
 
This article focuses on the results for the rotational analysis (\vsini\ and \vmacro) of the 285 likely single and SB1 stars of the total 415 galactic O stars in the IACOB and OWN sample, excluding SB2 systems and peculiar stars. This represents a step forward from previous studies in terms of the number of stars (+43\%), the separation between extra-broadening components, and the separation between different multiplicity statuses.
We made use of spectroscopic parameters obtained for the sample in previous articles (\Teff, \grav) and the multiplicity statuses to present a combined vision in the study of rotational properties.
The sample was divided into subsamples to investigate rotation on each of them, and their relationship in terms of evolution. The youngest subsample was studied in particular depth with the intention of exploring the \vini\ of the O-type stars. In each step we compared the empirical results with predictions by evolutionary models and population syntheses of O stars.

The main conclusions we reached with this work are:
\begin{itemize}
     \item The bimodal appearance of the general \vsini\ distribution persist with respect to previous works, as well as the absence of SB1 stars with extremely high rotational speeds.
     We support binary interaction, as depicted by the \cite{deMink2013} simulations, as the plausible origin of the bimodality feature, and the difficulty to detect binarity in the spun-up binary products, explained in \cite{deMink2014}, for the absence of very fast SB1 stars. All of this is in accordance with the \cite{Ramirez-Agudelo2013} results in the LMC.
     \item The extended width of the \vsini\ distribution of SB1 stars compared to the single distribution and its abrupt cutoff at 300 \kms\ agrees with tidal effects, since we could consider them stars without probable mass exchange interaction episodes \citep[][]{deMink2013}.
     \item We found an unexpected scarcity of very low-rotation stars ($<\sim$ 40-50 \kms) that appear to be concentrated on the lower part of the sHRD. 
     We claim microturbulence as the plausible origin, whose effects have not been completely eliminated, and which depends strongly on the location of the star in the sHRD \citep[see][]{Simon-Diaz2014}. 
     \item Stars with a very high rotation ($>$ 300 \kms) are grouped in the sHRD mostly below the 32 \msol\ evolutionary track.
     Their grouping in the sHRD may have its origin in the faster evolution in the higher-mass interaction products and/or mergers witnessed in the simulations of \cite{Wang2020}, and the increase in blue stragglers for smaller masses \citep[][]{Ahumada2007,Schneider2014,Schneider2015}.
     \item We do not detect a strong braking effect on the evolution of massive stars, even above the 32 \msol\ evolutionary track. This weak braking is in accordance with the Brott+11 models \citep{Brott2011}, instead of the strong and fast braking obtained in the Ekström+12 models \citep{Ekstroem2012}. 
     \item Using the \vini\ considered in most current single evolutionary models (\vini/\vcrit $\sim$ 0.4), the presence of stars with \vsini\ $>$300 \kms\ cannot be explained. 
     Binary interaction effects, as depicted in \cite{deMink2013}, could be a plausible explanation for the tail of the distribution and the slow rotation peak.
     
     \item We find that the empirical rotational properties of the ``young'' subsample of O-type stars is also close to a value \vini/\vcrit $\leq$ 0.2, in agreement with the trend for B stars presented in the literature \citep{Huang2010}. 
     Considering that the empirical distribution does not changes dramatically with evolution, this could be representative of the initial \vini/\vcrit.

\end{itemize}

In short, the empirical results lead us to believe that the majority of massive stars are only able to lose between 10-20\% of their rotation rate from birth. 
We consider that, in order to generate the observed distribution, the bulk of massive stars would need to be born with rotation rates closer to $\sim$150 \kms.
This corresponds to an initial rotation of $\sim$20\% of critical velocity.
These results match with the rotational study of O-type stars in the 30 Dor region of the LMC carried out by \cite{Ramirez-Agudelo2015}, with the interest of having a different metallicity, and our case being a study that comprises different clusters and field stars within the magnitude-limited sample of IACOB.
For a much more productive comparison with modeled values, new sets of models and/or population synthesis with an initial velocity \vini/\vcrit=0.2, and including binary interaction, should be generated. 

Collectively, our data represent a solid empirical basis to investigate the reaction and dependence of evolutionary models to changes in rotational properties, as well as other ingredients, such as mass loss or the inclusion of magnetic fields, as investigated in \cite{Keszthelyi2017} and  \cite{Keszthelyi2020}.
In the future, the general results we found with this work could be confirmed with an even more representative observed Galactic sample using data from the upcoming WEAVE survey \citep{WEAVE2012}.


\begin{acknowledgements}


\textit{In memoriam of Dr. Rodolfo H. Barba, whose dedication to the astrophysics was an inspiration for all.} 

We are extremely grateful to the referee of this paper, Zsolt Keszthelyi, for his valuable comments and for providing a constructive and helpful report to improve this work. 

G.H., S.S.-D. and A.H. acknowledge support from the Spanish Ministry of Science and Innovation (MICINN) through the Spanish State Research Agency through grants PGC-2018-0913741-B-C22 and the Severo Ochoa Programe 2020-2023 (CEX2019-000920-S). This work has also received financial support from the Canarian Agency for Economy, Knowledge, and Employment and the European Regional Development Fund (ERDF/EU), under grant with reference ProID2020010016. 
G.H. also acknowledge support from the Spanish Government Ministerio de Ciencia through grant PGC2018-95049-B-C22, and from the Science and Operations Department of the European Space Agency - Contract Number 4000126507-126507/19/ES/CM. 

Based on observations made with the Nordic Optical Telescope, operated by NOTSA, and the Mercator Telescope, operated by the Flemish Community, both at the Observatorio del Roque de los Muchachos (La Palma, Spain) of the Instituto de Astrofísica de Canarias.  

Based on observations at the European Southern Observatory in programs 073.D-0609(A), 077.B-0348(A), 079.D-0564(A), 079.D-0564(C), 081.D2008(A), 081.D-2008(B), 083.D-0589(A), 083.D-0589(B), 086.D-0997(A), 086.D-0997(B), 087.D-0946(A), 089.D-0975(A) 

This work has made use of data from the European Space Agency (ESA) mission
{\it Gaia} (\url{https://www.cosmos.esa.int/gaia}), processed by the {\it Gaia}
Data Processing and Analysis Consortium (DPAC,
\url{https://www.cosmos.esa.int/web/gaia/dpac/consortium}). Funding for the DPAC
has been provided by national institutions, in particular the institutions
participating in the {\it Gaia} Multilateral Agreement.

This paper made use of the IAC Supercomputing facility HTCondor (http://research.cs.wisc.edu/htcondor/), partly financed by the Ministry of Economy and Competitiveness with FEDER funds, code IACA13-3E-2493. 

G.H. wants to thank N. Langer and the Bonn group for their helpful comments that improved this manuscript significantly, and M. Cerviño for the precise corrections. 

\end{acknowledgements}


\bibliographystyle{aa}
\bibliography{ms}


\clearpage


\begin{appendix}

\section{Identification of SB1 stars in our working sample of O-type stars}\label{quan_var}

In this appendix, we briefly outline the strategy we followed to identify the stars in our working sample that can be tagged as SB1 to a high level of confidence\footnote{using our available spectroscopic data\,set}. A more detailed description about this process can be found in \cite{Holgado2019}, where we extend the criteria already presented in \cite{Holgado2018} to the case of stars with broad line-profiles (i.e., stars with \vsini\,$\gtrsim$\,150~\kms).


As a first step, we performed a visual inspection of a set of diagnostic lines which allowed us to optimize the identification of SB2 systems and to flag (qualitatively) the stars for which we had more than one epoch as presumably constant ($C$), line-profile variable ($LPV$), and clear or most likely SB1 systems. We refer the reader to \cite{Holgado2018} for some illustrative examples of the various types of detected variability.

As a second step, once we had access to the outcome of the {\sc iacob-broad} and {\sc iacob-gbat} analyses of those stars not identified as SB2s (see Sect.~\ref{sectionObsMeth}), we performed a (more) quantitative reassessment of the abovementioned classification.
In brief, the \fastwind\ synthetic spectrum of the {\sc iacob-gbat} best fitting model for each star (convolved with the corresponding \vsini, \vmacro, and $R$) was used to improve the radial velocity determination of each associated individual (multi-epoch) spectrum by means of a simple (iterative) cross-correlation technique. 

To this aim, we considered the subsequent initial set of lines included in the grid of \fastwind\ models presently incorporated to {\sc iacob-gbat}:  \ioni{He}{i}\,$\lambda$4026, \ioni{He}{i}\,$\lambda$4387, \ioni{He}{i}\,$\lambda$4471, \ioni{He}{i}\,$\lambda$4713, \ioni{He}{i}\,$\lambda$4922, \ioni{He}{i}\,$\lambda$5015, \ioni{He}{i}\,$\lambda$5047, \ioni{He}{i}\,$\lambda$5875, \ioni{He}{ii}\,$\lambda$4200, \ioni{He}{ii}\,$\lambda$4541, \ioni{He}{ii}\,$\lambda$4686, and \ioni{He}{ii}\,$\lambda$5411. Each line was analyzed separately, and the associated spectral window selected automatically using the corresponding synthetic line. The lines that were in emission or absent in the observed spectrum were automatically eliminated from the very beginning on a star-by-star basis. Then we followed an iterative process in which the lines providing \vrad\ measurements that deviated more than 2$\sigma$ from the mean were removed. Final values and associated uncertainties were then obtained from the mean and standard deviation of those individual \vrad\ measurements resulting from the remaining set of lines.

As a last step, we computed the dispersion -- $\sigma$(\vrad) -- and the peak-to-peak amplitude -- $\Delta$\vrad\ -- associated with each set of radial velocity measurements for a given star. Then, as described below, we used either of these two quantities to reassess the flags assigned to the targets initially identified as {\em C}, {\em LPV}, and SB1. In particular, we followed two different criteria depending on the projected rotational velocity of the star:

\begin{itemize}
 \item if \vsini\,$\leq$\,180~km\,s$^{-1}$, we flagged those cases with $\Delta$\vrad\ in the range  5\,--\,20~\kms\ and $\sigma$(\vrad) in the range 2.5\,--\,10~\kms\  as {\em LPV}, while those cases with lower (higher) dispersion were flagged as {\em C} (SB1), respectively;
 \item if \vsini\,$>$\,180~km\,s$^{-1}$, we flagged those cases with a ratio $\Delta$\vrad/\vsini\ in the range 0.02\,--\,0.1 and $\sigma$(\vrad)/\vsini\ in the range  0.01\,--\,0.05\ as {\em LPV}, while those cases with lower (higher) ratios were flagged as {\em C} (SB1), respectively.
\end{itemize}

If either of the two conditions was met, the object was reclassified. The indicator with the greatest variability always prevailed, that is, SB1, then LPV, and finally C.

This separation between stars with low and high projected rotational velocities accounts for the fact that subtle changes in the shape of the line profile due to, for example, pulsations, could led to a larger dispersion in the measured (multi-epoch) radial velocities  in stars with broad profiles. This situation could be erroneously interpreted as the detection of signatures associated with spectroscopic binarity, if not correctly taken into account.

\begin{figure}[h]
        \includegraphics[width=0.50\textwidth]{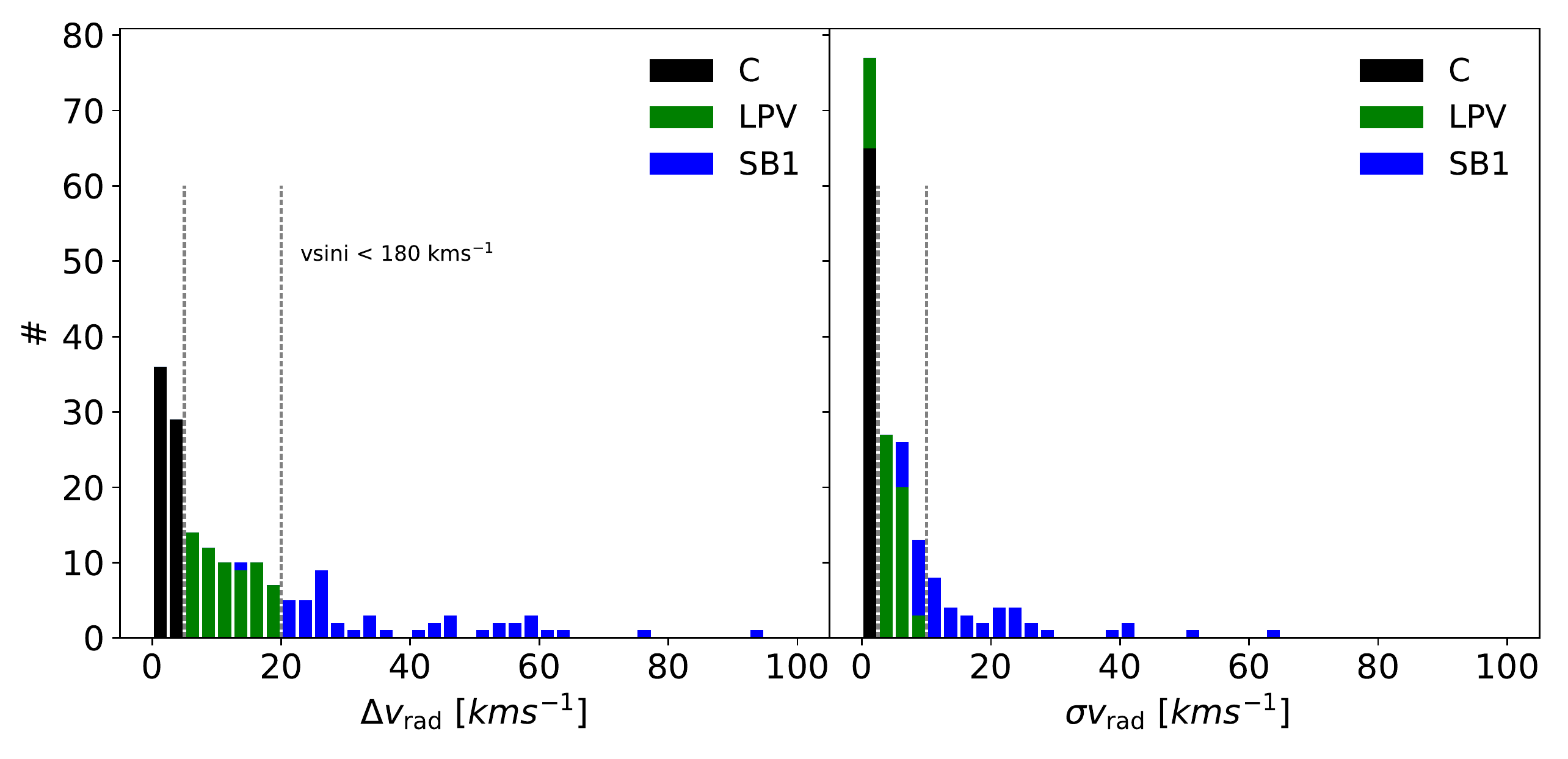}
        \caption{$\Delta$\vrad\ (left) and $\sigma$\vrad\ (right) distributions for the O-type stars in our initial sample not identified as SB2s and having a \vsini$\leq$\,180~\kms. We indicate with different colors how the stars were finally classified as $C$/$LPV$/SB1 in the stars with more than one spectrum. Horizontal dashed lines indicate the thresholds considered to separate the stars in the three categories. For most stars, both indicators coincide. In case of divergence, the higher variability flag prevails (SB1, then LPV, and then C).}
        \label{vsini_LPV_below} 
\end{figure}

\begin{figure}[h]
        \includegraphics[width=0.50\textwidth]{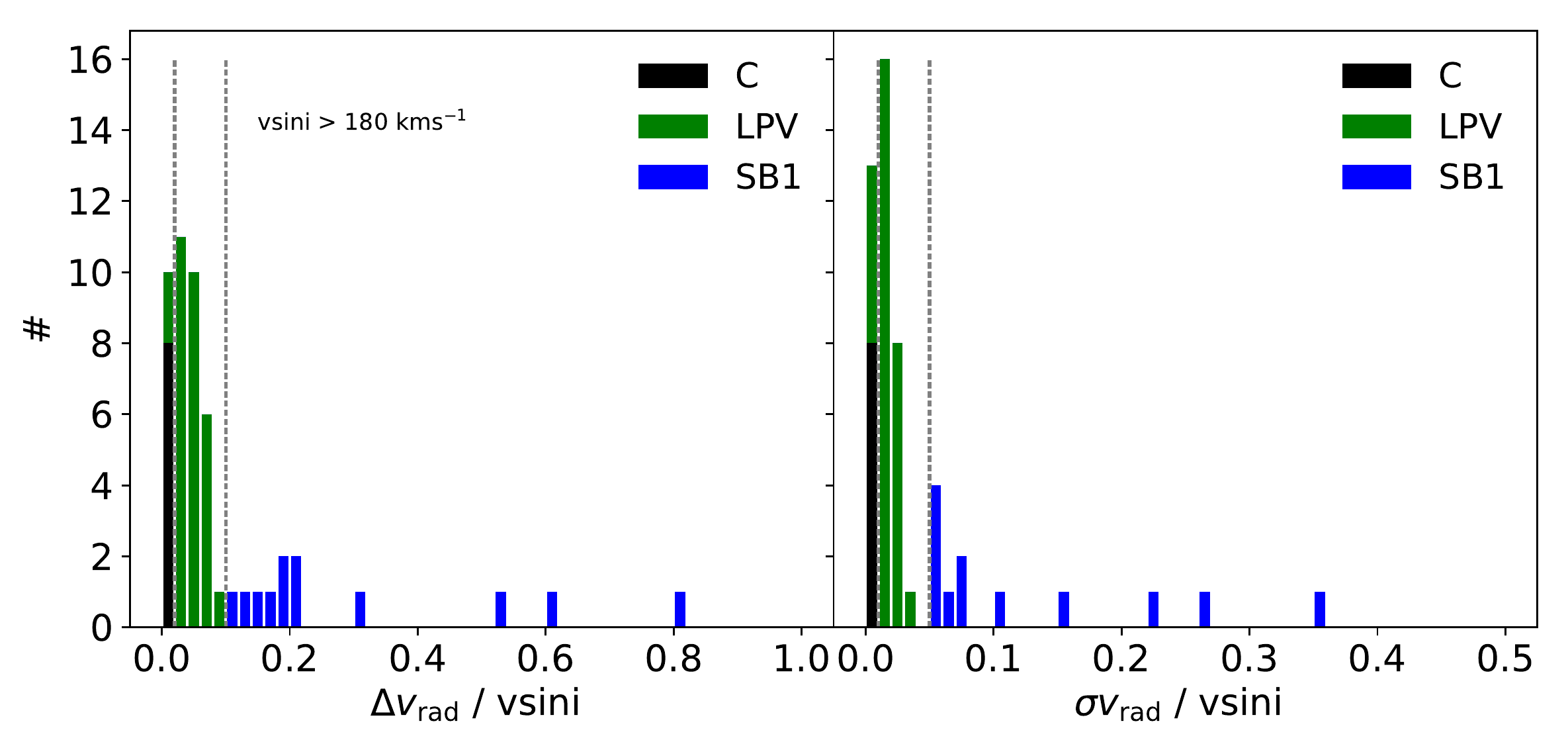}
        \caption{Similar to Fig.~\ref{vsini_LPV_below}, but for stars with a projected rotational velocity larger than 180~\kms. We note that, in this case, we consider the ratio of $\Delta$\vrad\ (left) and $\sigma$\vrad\ (right) with respect to \vsini.}
        \label{vsini_LPV}
\end{figure}

Figures~\ref{vsini_LPV_below} and \ref{vsini_LPV} present a summary of the outcome of this second assessment of spectroscopic variability in the targets initially flagged as {\em C}/{\em LPV}/SB1. The limiting values of $\Delta$\vrad\ and $\sigma$(\vrad) -- or their respective ratios with respect to \vsini\ -- quoted above are indicated as vertical dotted lines. 

\FloatBarrier

\FloatBarrier

\FloatBarrier

\section{Peculiar stars in the sample}\label{Peculiar}


\begin{figure}
\includegraphics[width=0.5\textwidth]{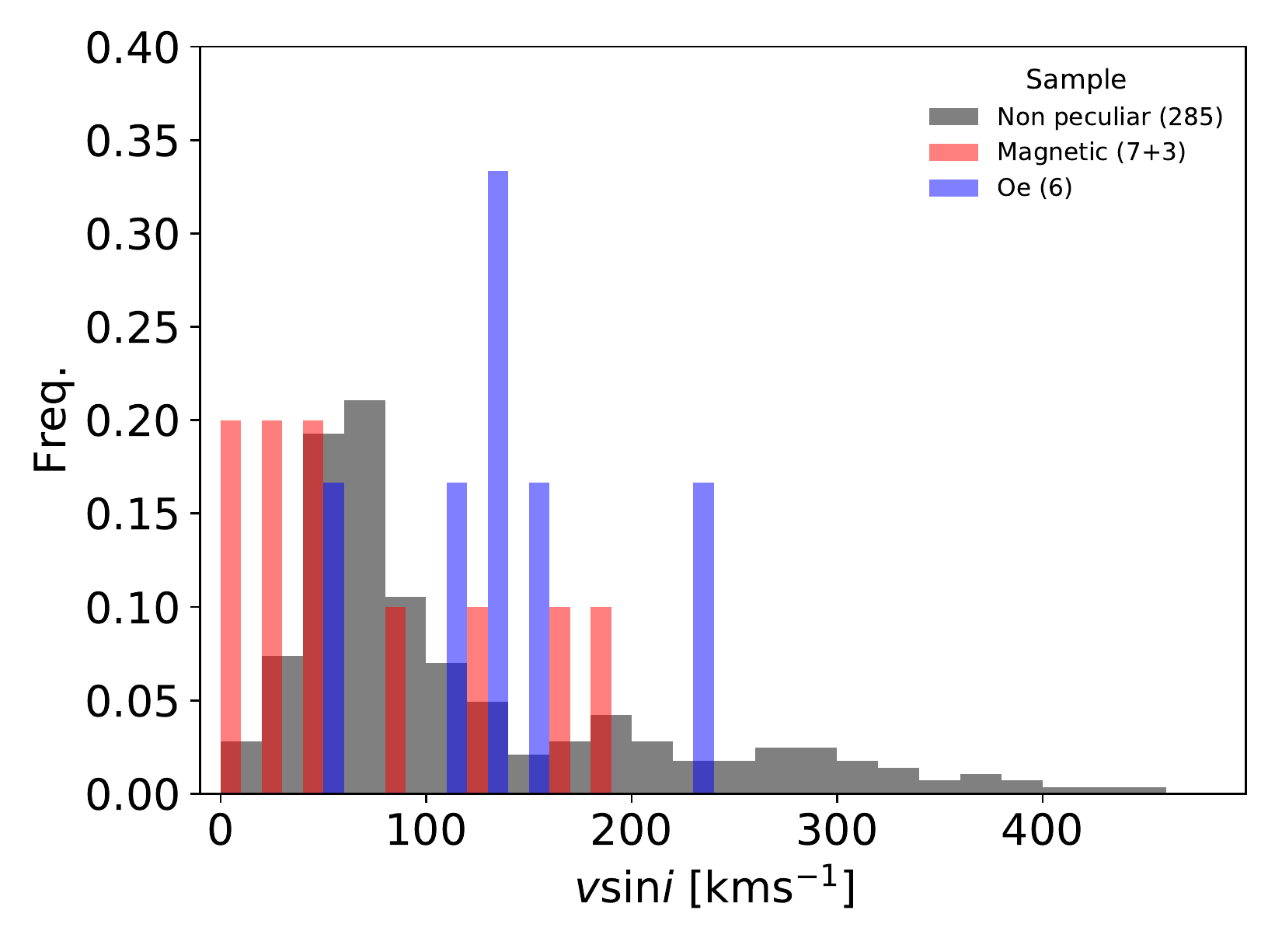}
\caption{Normalized \vsini\ distribution of the 13 peculiar stars in the sample (six Oe and seven(plus three) magnetic stars, see text. The global (also normalized) \vsini\ distribution of the full working sample of 285 LS and SB1 stars is depicted in the background for reference.}
\label{Peculiar_Hist}
\end{figure}

As commented in Sect.~\ref{sectionObsMeth}, there are 13 peculiar stars in our sample (six Oe and seven magnetic, see Table~\ref{tableValues_NoAnalisis_OeMag}) that were not further considered in our study because of the inherent limitations of our analysis strategy to provide reliable stellar parameters for these stars. However, we could obtain \vsini\ estimates for all of them. Figure~\ref{Peculiar_Hist} shows the \vsini\ distribution for this small sample of stars,where the global \vsini\ distribution of the remaining LS and SB1 stars is also depicted in the background for reference.

Magnetic stars are expected to present a distinct \vsini\ distribution when compared with the global sample \citep{ud-Doula2002, Grunhut2017}. Even though the coupling between magnetic fields and winds reduces mass loss, the presence of a magnetic field induces an extra braking effect an order of magnitude more efficient than wind-momentum loss \citep{ud-Doula2009,Petit2013}. 

For the Oe, as explained in \cite{Negueruela2004}, they represent the natural continuation of Be stars toward higher masses. There is a strong correlation between fast rotation and the Be phenomenon, manifested clearly when examining the $\Omega$/$\Omega_{\rm crit}$ factor, the surface angular velocity as a fraction of the critical breakup velocity. 

The number of magnetic and Oe stars within the sample is low, but Fig.~\ref{Peculiar_Hist} shows that most Oe stars have \vsini\ values above the main peak of the distribution ($>$100~\kms). Only one Oe star has a \vsini$<$50~\kms, and the origin could be in the inclination effect. 
For the magnetic stars, the majority present values within the main peak of the \vsini\ distribution, and only two have an unexpected high value of \vsini\ (\vsini$>$100~\kms). These are the companion stars of the black hole $Cyg$~$X$-1 and the post-interacting binary $\zeta$~$Pup$ \citep{Howarth2019}.
We note that the latter, $\zeta$~$Pup$, has a peculiar spectral type ("fp"), but the detection of its magnetic field has been considered spurious \citep[][]{Grunhut2017}.

In addition to these seven magnetic stars with problems in the spectroscopic fitting, the sample includes three other stars recognized as having a detectable magnetic field: HD~57682, HD~37742, and HD~54879 \citep[see e.g.,][]{Keszthelyi2019}. These three stars are included in the general sample as they do not exhibit any caveat in the automated \fastwind\ model fit, and they do not represent a very significant sample within the total.

\section{Comparison with \cite{Ramirez-Agudelo2013}}\label{RAGUAppend}

\begin{figure}[h]
        \centering
        \includegraphics[width=0.5\textwidth]{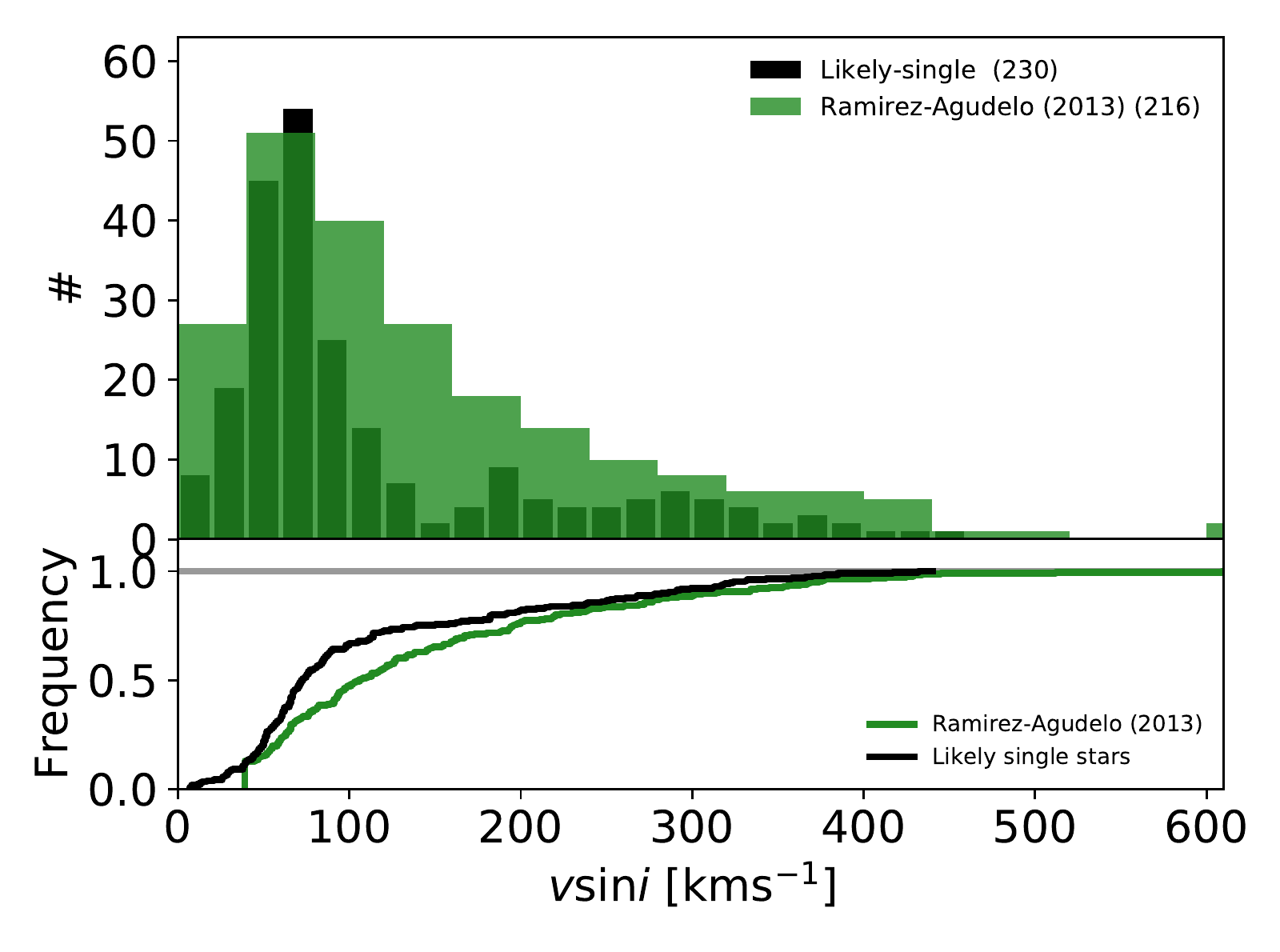}
\caption{\vsini\ histogram (up) and cumulative histogram (down) for \cite{Ramirez-Agudelo2013} and our sample. It includes the results for our sample of Galactic likely single O-type stars (black).}
        \label{hist_vsini_cumul_RAGU}
\end{figure}

Figure~\ref{hist_vsini_cumul_RAGU} presents the histogram and the associated cumulative histogram with the \vsini\ distribution obtained by \cite[][]{Ramirez-Agudelo2013} for a sample of 216 likely single stars O-type in the 30 Doradus region, of the LMC. This is a representative distribution of \vsini\ values for stars in a different metallicity environment (Z$_{\rm LMC}\sim0.5Z_{\odot}$). Their \vsini\ values were derived using the same methodology as our work, separating the macroturbulent broadening contamination. However, the lower resolving power (R$\sim$8000) and more limited wavelength coverage of their spectra imposed some caveats in the line-broadening characterization of their sample of stars. For example, their resolution implied that below 40 \kms\ any \vsini\ determination was uncertain. 
In addition, their reduced wavelength range force them to rely, in almost all cases, on \ioni{He}{i} and \ioni{He}{ii} lines (which are not optimal lines to determine \vsini\ in O-type stars, \citeauthor{Holgado2019} \citeyear{Holgado2019}). 

%

The main expected difference with respect to the Galactic case is that, at lower metallicities, the O-type stars are characterized by developing weaker winds \citep{Vink2000, Mokiem2006}. This reduces the amount of angular momentum that the star loses through the wind, which would finally produce a \vsini\ distribution with a higher average speed and a possible greater extreme rotation than in the Galactic case.

While the latter effect can be actually identified when comparing both distributions (the fast-rotators tail of the LMC distribution reach up\footnote{We note that this is not an effect of including SB2 stars, enhancing the number of fast rotators, as their sample only includes the likely single O-type stars.} to 600 \kms), no clear conclusions can be extracted in the low-velocity peak (below $\sim$40~\kms, or even $\sim$100~\kms\ in some cases). This is because of the limitations that affect the \vsini\ measurements of the LMC distribution.

\section{Understanding whether microturbulence is the culprit of the scarcity of low-\vsini\ stars in the O-star domain}\label{lowvalues}

\begin{figure}
\includegraphics[width=0.5\textwidth]{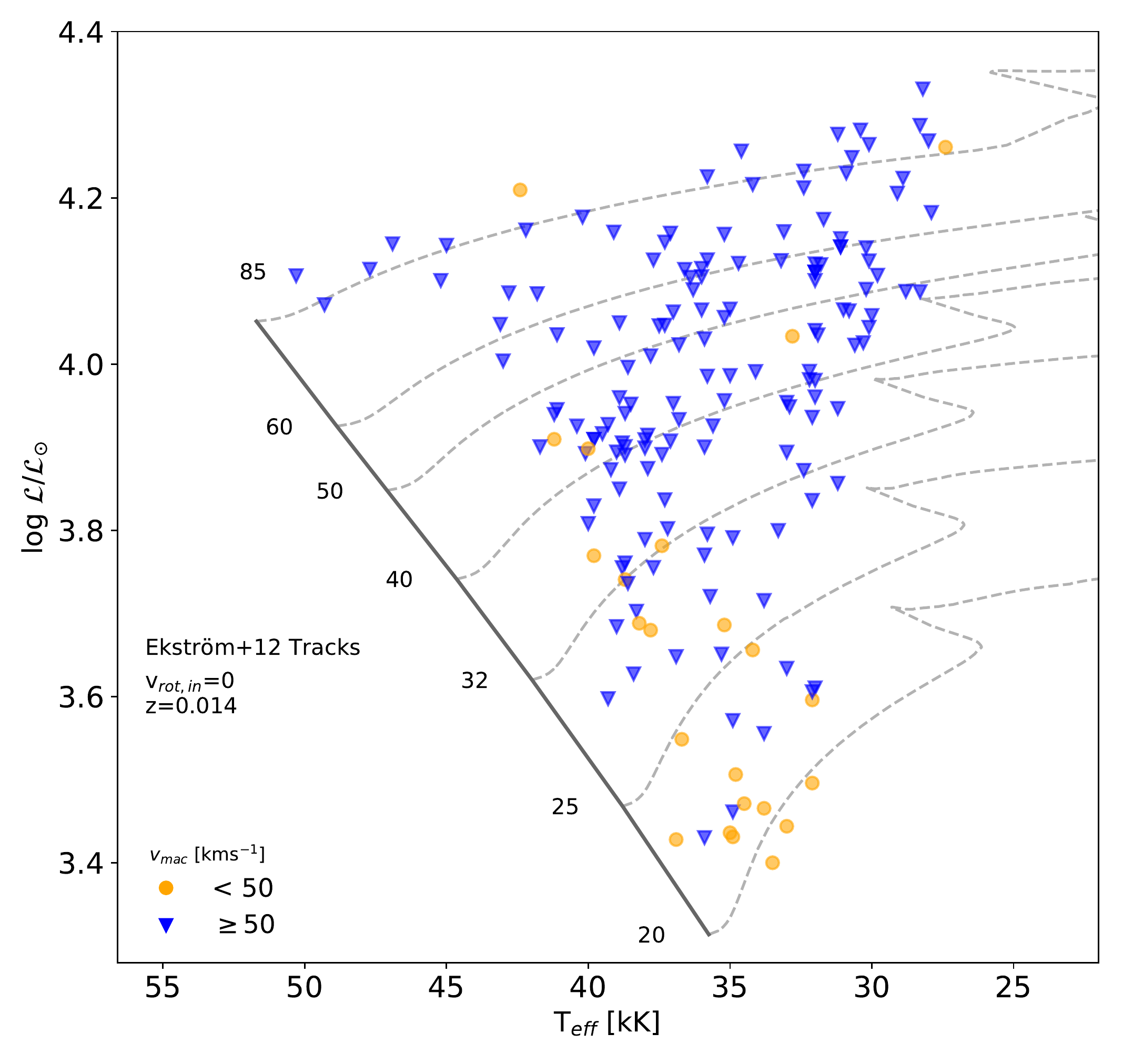}
\caption{Location in the sHRD of a subsample of 174 LS and SB1 stars with \vsini\,$<$\,100~\kms, separating the targets with a \vmacro\ lower or higher than 50~\kms.
}
\label{MacroOverRot}
\end{figure}

We have indicated on several occasions throughout this paper that there is increasing empirical evidence that our line-broadening analysis methodology might be failing under certain limiting circumstances in the low-\vsini\ regime \citep[see also][]{Sundqvist2013, Simon-Diaz2014, Markova2014}. For example, in the histogram presented in Fig.~\ref{Gen_Sample_Hist}, there is a notorious scarcity of O stars with spectroscopically inferred \vsini\ below $\sim$40-50~\kms\ (particularly in the case of stars with LCs, I, II and III). Also, panel A of Fig.~\ref{cuadrvsini} highlights that the only 21 stars in our working sample of 285 O-type stars with \vsini$<$35~\kms\ are concentrated below the 32~\msol\ tracks in the sHRD. Lastly, Fig.~\ref{logLsp_vs_Vsini} clearly shows that there exists a strong correlation between the lower limit in \vsini\ imposed by our analysis methodology and the quantity log~$\mathcal{L}$.

None of these results are expected if we assume a random distribution of inclination angles $i$ and we consider that more than 75\% of the investigated stars have a \vsini\,$<$\,150~\kms. Under these circumstances, one would roughly expect to find at least $\sim$15\% of the stars in the sample (independently of their location in the sHRD) with a \vsini\ lower than 35~\kms.  

Following ideas presented in \cite{Gray1973}, one of the possible culprits that might be at the origin of this problem is the effect of microturbulence. This affects the shape of the line-profiles and their associated FTs \citep[see notes in Sect.~\ref{lowvalues_intext} and][]{Simon-Diaz2014}.

While a thorough investigation of this subject is planned for a separate paper, we provide a few preliminary results of interest below, which can be obtained using the already available information (including rough estimations of the microturbulent velocity resulting from the {\sc iacob-gbat} analysis, not quoted in Table~\ref{tableValues}, and available at \citeauthor{Holgado2019} \citeyear{Holgado2019}).

The first result is summarized in Fig.~\ref{MacroOverRot}. With this figure we want to highlight that there seems to exist a correlation between the quantity \vmacro\ provided by {\sc iacob-broad} and the detectability of stars with \vsini\ below 35~\kms\ (see panel A in Fig.~\ref{cuadrvsini}). To better illustrate this statement, we only include in the figure the stars with a \vsini\ estimate below 100~\kms\ \citep[following][above this value, the \vmacro\ estimates are more uncertain and normally limited to upper limits]{Simon-Diaz2014}. We separate the sample in two groups, considering \vmacro\,=\,50~\kms\ as the limiting boundary. Interestingly, the region where stars for which a \vmacro\,$\>$50~\kms\ has been measured roughly match the region where no stars with \vsini\,$<$\,35~\kms\ are found.

This result could be interpreted as empirical evidence indicating that we are still not adequately disentangling  the contribution of the so-called macroturbulent broadening from the pure effect of rotation. Alternatively, it could indicate that there are other sources of non-rotational line-broadening that are still affecting our \vsini\ estimations,  microturbulence being one of the possible sources. Indeed, taking into account that, by means of the way we are using {\sc iacob-broad}\footnote{We consider that the intrinsic profile, which is later on convolved with the instrumental+rotational+macroturbulent profiles during the analysis process, is a $\delta$-function with the same equivalent width as the observed line-profile.}, the quantity \vmacro\ resulting from the GOF analysis encloses any source of non-rotational extra-broadening, a large value of \vmacro\ might also indicate that the line-profile is affected by a relatively high microturbulence. If this is the case, the larger the microturbulence, the lower the frequency (or, equivalently, the higher the "equivalent" \vsini) at which the associated zero will be found in the FT.

%
\begin{figure}[!h]
\includegraphics[width=0.5\textwidth]{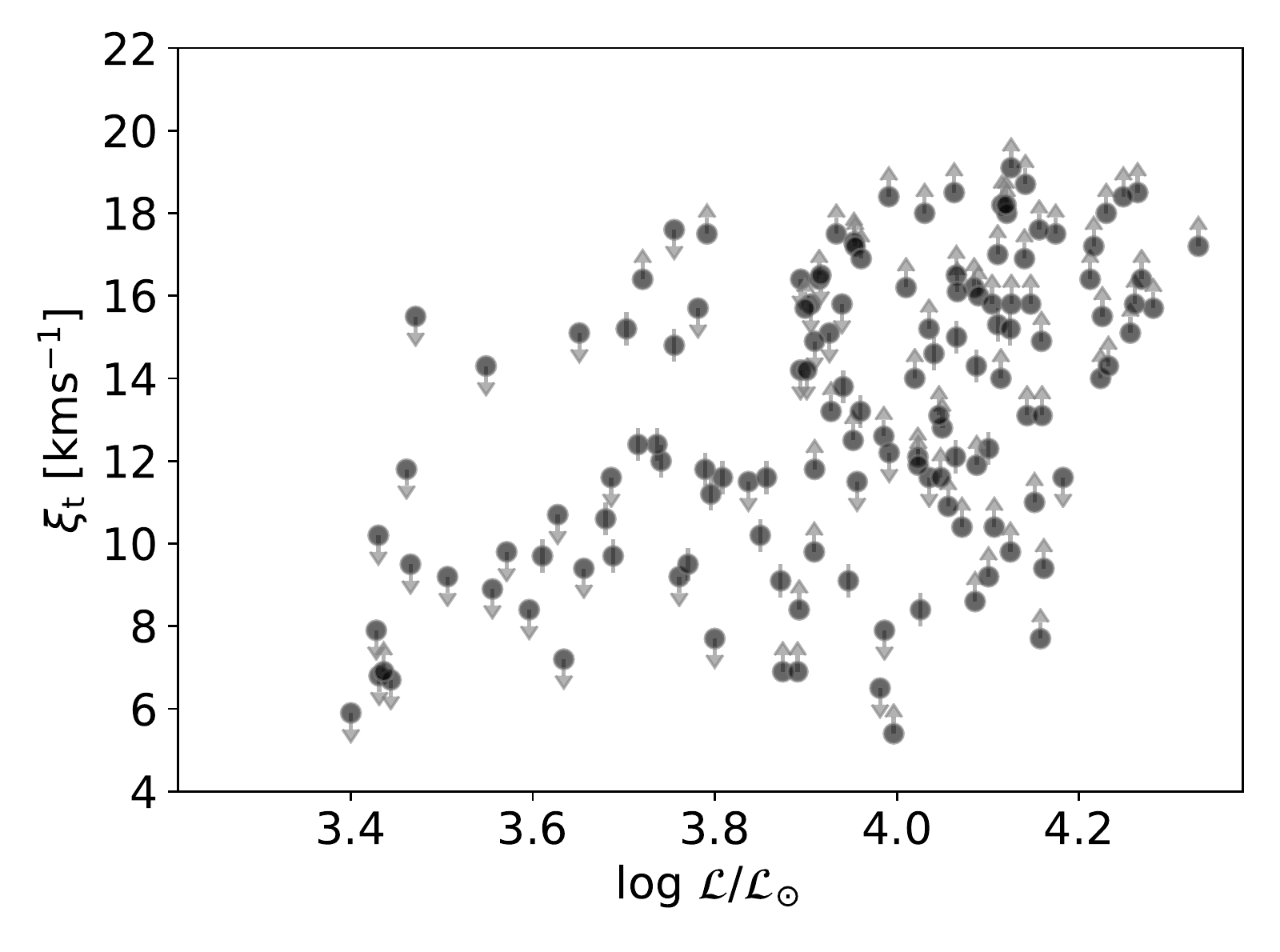}
\caption{\micro\,--\,\Lsp\ diagram including only those stars in our working sample with \vsini\,$<$\,80~\kms\ and \micro\ available. 
}
\label{Vmic_LogLsp}
\end{figure}

As commented in \cite{Holgado2018}, a proper characterization of microturbulence in the O-star domain is still lacking due to the more limited number of metal lines available to determine accurate values of this parameter (e.g., using the curve of growth method), compared to the case of B-type stars. Despite this limitation, several studies have found suggestive evidence of the existence of a correlation between microturbulence -- obtained from the analysis of the \ion{He}{i} lines -- and luminosity class in these type of stars \citep[e.g.,][]{Massey2013, Markova2014, Holgado2018}.

Keeping in mind that the accuracy reached in the estimation of the microturbulent velocity (\micro) is not so good when using He lines as when metal lines are considered, we decided to use the outcome of {\sc iacob-gbat} to investigate whether we find any interesting correlation between \micro\ and \Lsp. Figure~\ref{Vmic_LogLsp} summarizes the result of this investigation, where we only consider those stars in our working sample having a \vsini\,$<$\,80~\kms. This sort of correlation seems to be present, although with a significant scatter.

%
\begin{figure}[!h]
\includegraphics[width=0.5\textwidth]{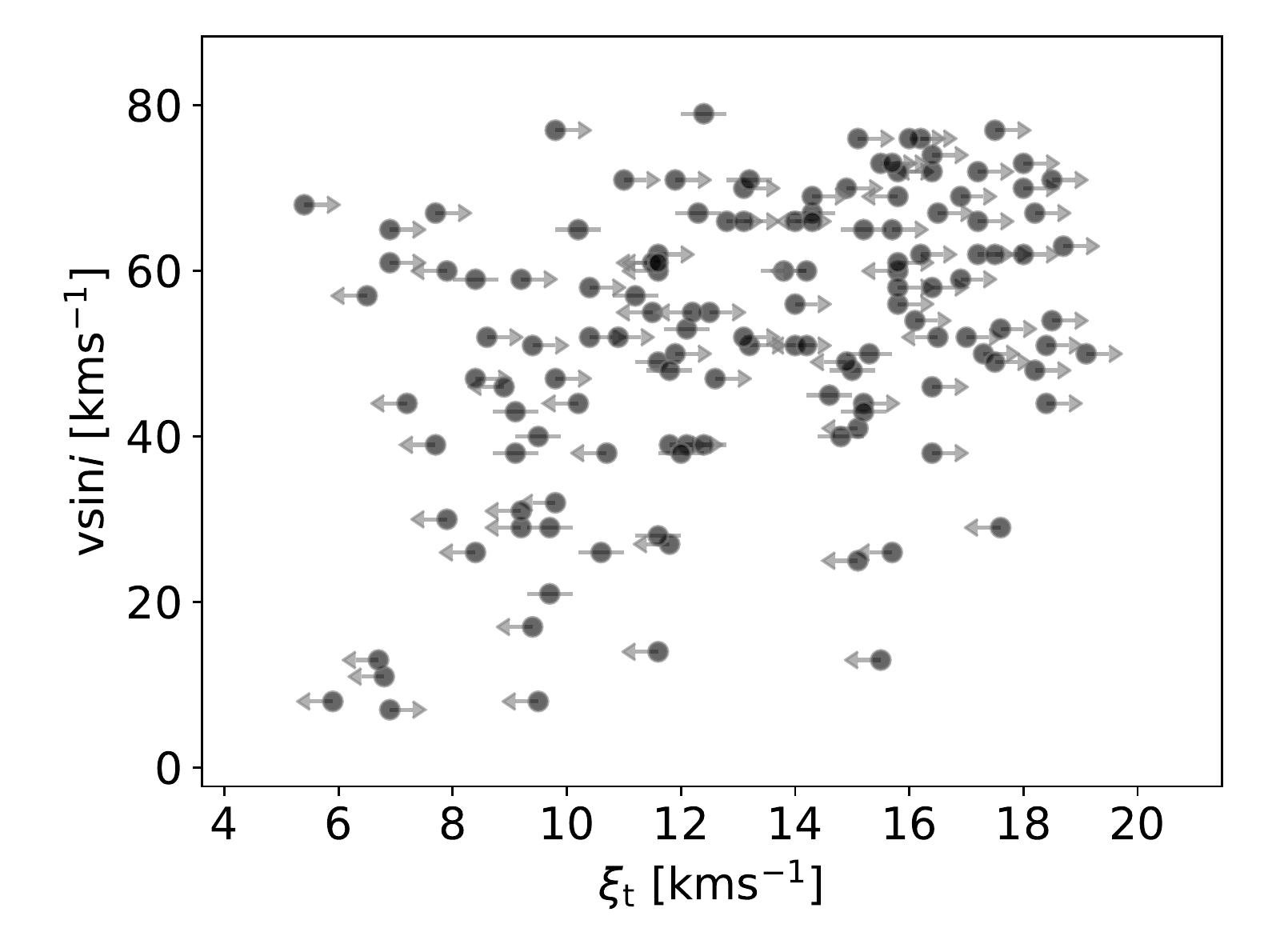}
\caption{Distribution in the \vsini\,--\,\micro\ diagram of the same set of stars considered in Fig.~\ref{Vmic_LogLsp} .}
\label{Vmic_vsini_VH}
\end{figure}

Awaiting a more accurate determination of \micro\ in the stars depicted in Fig.~\ref{Vmic_LogLsp}, we performed the following formal exercise to determine the location in the FT (in terms of "equivalent" \vsini) of the zero associated with a given value of \micro\ covering the range shown in Fig.~\ref{Vmic_LogLsp}. To this aim, we calculated a grid of {\sc fastwind} models including H, He, and O as explicit elements. The surface gravity was fixed to \grav\,=\,3.8 dex, and we covered a range in \Teff\ from 25 to 50 kK. For each of the models, we computed a synthetic spectrum assuming two values of \micro\ in the formal solution (5 and 20~\kms), and each spectrum was convolved to a resolving power $R$=50000 and a \vsini\,=\,5~\kms. We then used {\sc iacob-broad} to perform the analysis of all the \ion{O}{iii}$\lambda$5592\AA\ synthetic lines resulting from the {\sc fastwind} computations. Those cases in which the estimated \vsini(FT) was larger than 5~\kms\ indicated that the first zero of the FT was actually associated with microturbulence and not with rotation \cite[see also the outcome of a similar exercise in][]{Simon-Diaz2014}.

Interestingly, we found that, while in the case of the \ion{O}{iii} synthetic lines with \micro\,=\,5~\kms, the values of \vsini(FT) that were recovered were close to 10~\kms, much larger values (up to 30~\kms) were obtained when considering the synthetic profiles computed with a \micro\,=\,20~\kms. In addition, we found that the resulting \vsini(GOF) estimates provided by {\sc iacob-broad} (assuming a $\delta$-function as intrinsic profile) were very similar to those corresponding to the erroneously estimated \vsini(FT) values. Therefore, this simple academic exercise served to illustrate that classical microturbulence could explain both the observed tendency in Fig.~\ref{logLsp_vs_Vsini}, as well as the good agreement found between \vsini(FT) and \vsini(GOF) in Fig.~\ref{GOF_FT} in the low- \vsini\ range, even for cases in which \vsini\ is most likely affected by methodological limitations.

To reinforce this statement, in Fig.~\ref{Vmic_vsini_VH} we present again the distribution of LS and SB1 stars with \vsini\,$<$\,80~\kms, but this time in the \vsini\,--\,\micro\ diagram. Roughly speaking, there seems to be an increasing tendency in the lower \vsini\,, which is achieved for increasing values of \micro.

In summary, although there have been advances in disentangling the pure effect of rotation from other sources of line-broadening to get more reliable estimates in \vsini\,, we are still not getting the correct rotational measurements for very low values of \vsini. The inadequately disentangled effect, which is most likely produced by microturbulence, seems to be hampering our ability to reach actual \vsini\ estimates below $\sim$40\,--\,50~\kms. We thus concur with the precautions raised by \cite{Sundqvist2013} regarding the interpretation of any statistical distribution of observed rotation rates of massive-star populations. However, this effect most likely does not affect our measurements above $\sim$50\,--\,60~\kms\ or the measurement of \vsini\ in stars with low microturbulent velocities.

\section{Impact of bin size} \label{AppBines}

\begin{figure}[!h]
\includegraphics[width=0.5\textwidth]{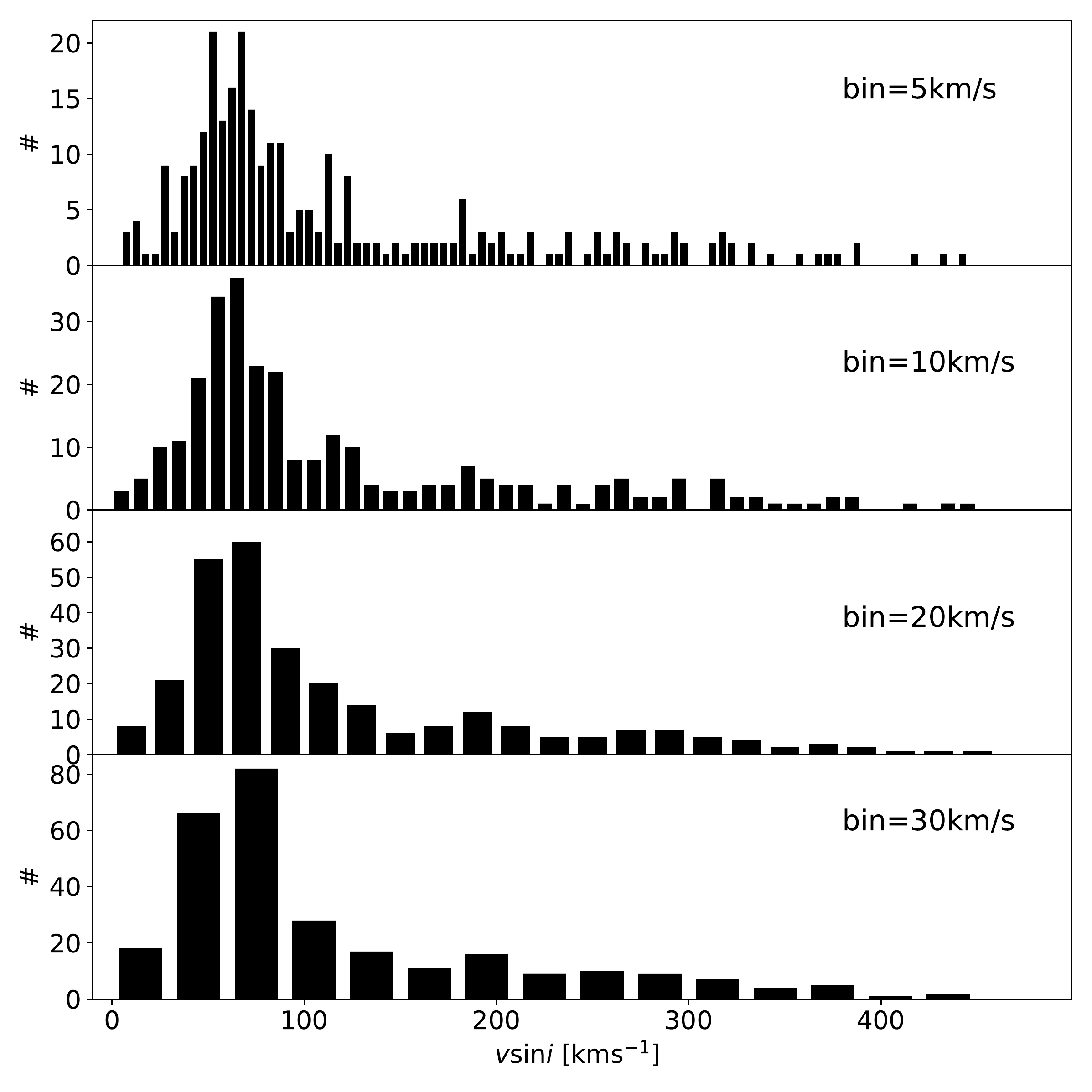}
\caption{\vsini\ distribution for the whole sample using different bin sizes.}
\label{Bines}
\end{figure}

The bin size used throughout this study for the \vsini\ histograms is 20\kms\,, following the reasoning presented in Sect.~\ref{GenOv}. Figure~\ref{Bines} shows the effect of selecting different bin sizes, and how distributions could be misinterpreted as a result.

\FloatBarrier
%
\FloatBarrier

\section{Tables}
\onecolumn

{
\noindent
\begin{longtable}{l@{\hskip 0.1in}r@{\hskip 0.05in}l@{\hskip 0.1in}ccc@{\hskip 0.1in}c@{\hskip 0.1in}c@{\hskip 0.1in}c@{\hskip 0.1in}c@{\hskip 0.1in}cl}
\caption{List of the 285 stars with spectroscopic parameters obtained in this work. Stars are separated by luminosity class (horizontal lines) and sorted by spectral type. Columns are: (1, 2) Name and spectral type from GOSC, (3, 4, 5) \Teff, \gravt\ and log~$\mathcal{L}$ \citep[from][]{Holgado2020}. From this work are (6) diagnostic line used to obtain the line-broadening parameters (\vsini, \vmacro) quoted in columns (7, 8). (9) is number of epochs available. (10) marks the evolutionary subsample assignment following Fig.~\ref{Mezcla_histogr}, and (11) whether the star was clearly identified as LPV/SB1 from the available multi-epoch observations. Typical uncertainties are $\Delta$\Teff=0.5 kK, $\Delta$\gravt=0.1 dex, $\Delta$\Lsp=0.12 dex, and 10\% for \vsini\ and \vmacro.}\\
\label{tableValues}\\ 
\hline \hline
\noalign{\smallskip}
Name &   SpT &             LC &      $T_{\rm eff}$ & \gravt & log $\mathcal{L}$/$\mathcal{L}$$_{\odot}$ & Line &  \vsini\  & \vmacro\ & Epochs & Evol. & Notes \\
     &       &                &           [kK]     & [dex] &            [dex]                        &  &   [\kms]        & [\kms] &  & sample & \\
\hline
\noalign{\smallskip}
\endfirsthead
\caption{continued.}\\
\hline
\hline
\noalign{\smallskip}
Name &   SpT &             LC &      $T_{\rm eff}$ & \gravt & log $\mathcal{L}$/$\mathcal{L}$$_{\odot}$ & Line & \vsini\  & \vmacro\ & Epochs & Evol. & Notes \\
     &       &                &           [kK]     & [dex] &            [dex]                        &  &   [\kms]        & [\kms] &  & sample & \\
\hline
\noalign{\smallskip}
\endhead
\noalign{\smallskip}
\noalign{\smallskip}
\hline\hline
\noalign{\smallskip}
\multicolumn{12}{l}{\textbf{Notes}:  LPV: Line profile variability or plausible SB1; SB1: Confirmed SB1. See appendix~\ref{quan_var}.} \\
\multicolumn{12}{l}{$\mathcal{L}=T_{\rm eff}^4$\,/$g_{\rm corr}$,\, $g_{\rm corr}=g_{\rm sp}+g_{\rm cent}$, $g_{\rm cent}\approx\frac{(V_{\rm rot}\sin i)^2}{R_{*}} $}\\
\multicolumn{12}{l}{\ioni{O}{iii}: \ioni{O}{iii}$\lambda$5592,
\ioni{Si}{iii}: \ioni{Si}{iii}$\lambda$4452,
\ioni{He}{i}: \ioni{He}{i}$\lambda$4387/4471/4713/5875,
\ioni{N}{v}: \ioni{N}{v}$\lambda$4603/4620,
\ioni{N}{iv}: \ioni{N}{iv}$\lambda$6380
}\\
\endfoot
    HD\,93129\,AaAb &    O2 &          I\,f* &   45.6 &   3.89 &   4.14 &    \ioni{N}{v} &            116 &       142 &  12  &   a  &  SB1 \\
         CYG\,OB2-7 &    O3 &          I\,f* &   50.3 &   4.09 &   4.11 &    \ioni{N}{v} &             75 &       126 &  1   &   a  &    . \\
         ALS\,15210 &  O3.5 &      I\,f*\_Nwk &  42.4 &   3.69 &   4.21 &    \ioni{N}{v} &             98 &         4 &  1   &   b  &    . \\
      HD\,190429\,A &    O4 &           I\,f &   36.6 &   3.53 &   4.11 &    \ioni{N}{v} &             90 &       113 &  3   &   b  &    . \\
          HD\,15570 &    O4 &           I\,f &   40.2 &   3.63 &   4.18 &    \ioni{N}{v} &             81 &       115 &  3   &   b  &    . \\
          HD\,16691 &    O4 &           I\,f & $>$39.7 &  $>$3.74 &  $<$4.05 & \ioni{N}{v} &          158 &        88 &  3   &   a  &  SB1 \\
          HD\,14947 &  O4.5 &           I\,f &   39.1 &   3.65 &   4.11 &  \ioni{O}{iii} &            114 &        22 &  3   &   b  &  LPV \\
         CYG\,OB2-9 &  O4.5 &           I\,f &   40.1 &   3.92 &   3.88 &   \ioni{He}{i} &            112 &        82 &  1   &   a  &    . \\
\end{longtable}
}

{

\noindent\begin{longtable}{l@{\hskip 0.1in}r@{\hskip 0.05in}l@{\hskip 0.1in}c@{\hskip 0.1in}c@{\hskip 0.1in}c@{\hskip 0.1in}c@{\hskip 0.1in}cl}
\caption{\tiny List of the 19 stars with \vsini\ values $>$300\kms. Stars are separated by luminosity class (horizontal lines) and sorted by spectral type. 
}\\
\label{tableValues_over300}\\
\hline \hline
\noalign{\smallskip}
Name &   SpT &             LC &      $T_{\rm eff}$ &  log $\mathcal{L}$/$\mathcal{L}$$_{\odot}$ &   \vsini\  & \vmacro\ & Epochs &  Notes \\
     &       &                &           [kK]     &            [dex]                        &    [\kms]        & [\kms] &    \\
\hline
\noalign{\smallskip}
\endfirsthead
\caption{continued.}\\
\hline
\hline
\noalign{\smallskip}
Name &   SpT &             LC &      $T_{\rm eff}$ &  log $\mathcal{L}$/$\mathcal{L}$$_{\odot}$ &   \vsini\  & \vmacro\ & Epochs &  Notes \\
     &       &                &           [kK]     &            [dex]                        &  [\kms]        & [\kms] &    \\
\hline
\noalign{\smallskip}
\endhead
\noalign{\smallskip}
\noalign{\smallskip}
\hline\hline
\noalign{\smallskip}
\multicolumn{9}{l}{\tiny \textbf{Notes}:  LPV: Line profile variability or plausible SB1; SB1: Confirmed SB1. See appendix~\ref{quan_var}.} \\
\multicolumn{9}{l}{$\mathcal{L}=T_{\rm eff}^4$\,/$g_{\rm corr}$,\, $g_{\rm corr}=g_{\rm sp}+g_{\rm cent}$, $g_{\rm cent}\approx\frac{(V_{\rm rot}\sin i)^2}{R_{*}} $}\\
\endfoot
\noalign{\smallskip}
\noalign{\smallskip}
\hline\hline
\multicolumn{9}{l}{\tiny \textbf{Notes}:  LPV: Line profile variability or plausible SB1; SB1: Confirmed SB1. See appendix~\ref{quan_var}.} \\
\multicolumn{9}{l}{$\mathcal{L}=T_{\rm eff}^4$\,/$g_{\rm corr}$,\, $g_{\rm corr}=g_{\rm sp}+g_{\rm cent}$, $g_{\rm cent}\approx\frac{(V_{\rm rot}\sin i)^2}{R_{*}} $}\\
\noalign{\smallskip}
\multicolumn{9}{l}{
\begin{minipage}{40em}
\tiny 
  Columns are: (1, 2) Name and spectral type from GOSC, (3, 4) \Teff, and log~$\mathcal{L}$ \citep[from][]{Holgado2020}. From this work are  \vsini\ and \vmacro\ quoted in Columns 5 and 6. (7) is the number of epochs available, and (8) whether the star was clearly identified as LPV/SB1 from the available multi-epoch observations.
\end{minipage}}
\\
\endlastfoot
    HD\,191423 &   ON9 &     II-III\,nn &   32.3$\pm$0.8 &   3.72$\pm$0.07 &          432 &        ... &    7 &        LPV \\
 \hline
\noalign{\smallskip}     
     HD\,14442 &    O5 &     ...\,n(f)p &   39.1$\pm$1.3 &   4.14$\pm$0.14 &          324 &        ... &    7 &        LPV \\
    HD\,203064 &  O7.5 &    III\,n((f)) &   35.3$\pm$0.5 &   3.89$\pm$0.04 &          312 &        ... &   31 &        LPV \\
     HD\,93521 &  O9.5 &        III\,nn &   31.7$\pm$0.8 &   3.61$\pm$0.09 &          385 &        ... &   36 &        LPV \\
    HD\,117490 & ON9.5 &        III\,nn &   31.6$\pm$0.7 &   3.65$\pm$0.07 &          343 &        ... &   10 &        LPV \\
 \hline
\noalign{\smallskip}   
    HD\,229232 &    O4 &     V:\,n((f)) &   42.9$\pm$2.2 &   4.11$\pm$0.13 &          313 &        ... &    5 &        LPV \\
     HD\,14434 &  O5.5 &     IV\,nn(f)p &   38.6$\pm$1.1 &   3.78$\pm$0.12 &          417 &        ... &    6 &        LPV \\
    HD\,228841 &  O6.5 &      V\,n((f)) &   37.7$\pm$1.4 &   3.82$\pm$0.15 &          317 &        ... &    9 &        LPV \\
\end{longtable}
   }

{
\noindent\begin{longtable}{l@{\hskip 0.1in}r@{\hskip 0.05in}l@{\hskip 0.1in}c@{\hskip 0.1in}c@{\hskip 0.1in}c@{\hskip 0.1in}c@{\hskip 0.1in}cl}
\caption{\tiny List of the seven stars with \Teff\ values $\geq$45kK. Stars are sorted by spectral type. 
}\\
\label{tableValues_over45K}\\
\hline \hline
\noalign{\smallskip}
Name &   SpT &             LC &      $T_{\rm eff}$ &  log $\mathcal{L}$/$\mathcal{L}$$_{\odot}$ &   \vsini\  & \vmacro\ & Epochs &  Notes \\
     &       &                &           [kK]     &            [dex]                        &    [\kms]        & [\kms] &    \\
\hline
\noalign{\smallskip}
\endfirsthead
\caption{continued.}\\
\hline
\hline
\noalign{\smallskip}
Name &   SpT &             LC &      $T_{\rm eff}$ &  log $\mathcal{L}$/$\mathcal{L}$$_{\odot}$ &   \vsini\  & \vmacro\ & Epochs &  Notes \\
     &       &                &           [kK]     &            [dex]                        &    [\kms]        & [\kms] &    \\
\hline
\noalign{\smallskip}
\endhead
\noalign{\smallskip}
\noalign{\smallskip}
\hline\hline
\noalign{\smallskip}
\multicolumn{9}{l}{\tiny \textbf{Notes}:  LPV: Line profile variability or plausible SB1; SB1: Confirmed SB1. See appendix~\ref{quan_var}.} \\
\multicolumn{9}{l}{$\mathcal{L}=T_{\rm eff}^4$\,/$g_{\rm corr}$,\, $g_{\rm corr}=g_{\rm sp}+g_{\rm cent}$, $g_{\rm cent}\approx\frac{(V_{\rm rot}\sin i)^2}{R_{*}} $}\\
\noalign{\smallskip}
\multicolumn{9}{l}{
\begin{minipage}{40em}
\tiny 
  Columns are: (1, 2) Name and spectral type from GOSC, (3, 4) \Teff, and log~$\mathcal{L}$ \citep[from][]{Holgado2020}. From this work are  \vsini\ and \vmacro\ quoted in Columns 5 and 6. (7) is the number of epochs available, and (8) whether the star was clearly identified as SB1 from the available multi-epoch observations.
\end{minipage}}
\\
\endfoot
   HD\,93129\,AaAb &   O2 &      I\,f* &   45.6$\pm$1.1 &   4.14$\pm$0.11 &       116 &       142 &     12 &   SB1 \\
        CYG\,OB2-7 &   O3 &      I\,f* &   50.3$\pm$1.8 &   4.11$\pm$0.14 &        75 &       126 &      1 &    .  \\
         HD\,64568 &   O3 & V\,((f*))z &   46.9$\pm$1.0 &   4.14$\pm$0.07 &        75 &        53 &      3 &    .  \\
         HD\,93128 & O3.5 & V\,((fc))z &   49.3$\pm$2.2 &   4.07$\pm$0.18 &        58 &        56 &      3 &    .  \\
      HD\,93129\,B & O3.5 &  V\,((f))z &   47.7$\pm$1.8 &   4.11$\pm$0.10 &        66 &        62 &      1 &    .  \\
     HD\,93250\,AB &   O4 &   IV\,(fc) &   45.0$\pm$0.6 &   4.14$\pm$0.06 &        70 &        84 &      4 &    .  \\
         HD\,96715 &   O4 &  V\,((f))z &   45.2$\pm$1.2 &   4.10$\pm$0.13 &        59 &        86 &      2 &    .  \\
\end{longtable}
   }

{
\setlength\LTleft{1.3cm}  
\noindent\begin{longtable}{l@{\hskip 0.1in}r@{\hskip 0.05in}l@{\hskip 0.1in}c@{\hskip 0.1in}c@{\hskip 0.1in}c@{\hskip 0.1in}cl}
\caption{\tiny 13 peculiar stars (Oe and Magnetic) in the sample with no spectroscopic analysis available but line-broadening analysis completed. 
}\\
\label{tableValues_NoAnalisis_OeMag}\\
\hline \hline
\noalign{\smallskip}
Name &   SpT &             LC & Line & \vsini\  & \vmacro\ & Epochs & Notes \\
     &       &                &      &    [\kms]        & [\kms] & &  \\
\hline
\noalign{\smallskip}
\endfirsthead
\caption{continued.}\\
\hline
\hline
\noalign{\smallskip}
Name &   SpT &             LC & Line &  \vsini\  & \vmacro\ & Epochs & Notes \\
     &       &                &      &    [\kms]        & [\kms] &  & \\
\hline
\noalign{\smallskip}
\endhead
\noalign{\smallskip}
\noalign{\smallskip}
\hline\hline
\noalign{\smallskip}
\multicolumn{8}{l}{\tiny \textbf{Notes}:  LPV: Line profile variability or plausible SB1; SB1: Confirmed SB1. See appendix~\ref{quan_var}.} \\
\noalign{\smallskip}
\multicolumn{8}{l}{
\begin{minipage}{40em}
\tiny 
   Columns are: (1, 2) Name and spectral type from GOSC, (3) diagnostic line for line-broadening, (4) projected rotational velocity from the model fit, (5) macroturbulence, (6) epochs and (7) additional notes. Magnetic stars include reference to its magnetic detection. Mag: Magnetic. WR: Wolf-Rayet.
\end{minipage}}
\\
\multicolumn{8}{l}{
\begin{minipage}{40em}
\tiny 
   \textbf{References}: $^{(a)}$Spurious detection in \cite{Grunhut2017} and bad fitting, $^{(b)}$\cite{Hubrig2008}, $^{(c)}$\cite{Hubrig2011},$^{(d)}$\cite{Naze2008},$^{(e)}$\cite{Martins2010},$^{(f)}$\cite{Wade2011},$^{(g)}$\cite{Caballero-Nieves2009}.
\end{minipage}}
\\
\endfoot

   HD\,66811 &    O4 &      I\,(n)fp &    \ioni{N}{v} &         198 &        12 &    7 &      Mag$^{(a)}$ \\
  HD\,148937 &    O6 &      ...\,f?p &  \ioni{O}{iii} &          39 &        76 &   22 &      Mag$^{(b)}$ \\
  CPD\,-28\,2561 &  O6.5 &    .\,f?p &  \ioni{O}{iii} &         169 &        94 &   12 &    Mag$^{(c)}$ \\  
   HD\,37022 &    O7 &          V\,p &  \ioni{O}{iii} &          23 &        68 &    8 &Mag$^{(d)}$      LPV \\
     HD\,108 &    O8 &   ...\,fp\_var &  \ioni{O}{iii} &          46 &        86 &   4 &      Mag$^{(e)}$, SB1 \\
  HD\,191612 &    O8 &  ...\,f?p\_var &  \ioni{O}{iii} &          45 &        60 &   2 & Mag$^{(f)}$,      LPV \\
  HD\,226868 &  O9.7 &    Iab\,p\_var &  \ioni{O}{iii} &          95 &        70 &   5 &      Mag$^{(g)}$, SB1 \\
\hline
\noalign{\smallskip}
   HD\,39680 &    O6 & V:\,[n]pe\_var &   \ioni{He}{i} &         101 &        31 &   4 &       Oe, SB1 \\
\end{longtable}
   }

\end{appendix}
        
\end{document}